\documentclass[aps,prl,preprint,superscriptaddress]{revtex4-2}
\usepackage{physics}
\usepackage{graphicx}
\usepackage{siunitx}
\usepackage{makecell}
\usepackage{xcolor}
\usepackage{latexsym}
\usepackage{graphics}
\usepackage{amsmath}
\usepackage{tikz}
\usepackage{amsfonts}
\newcommand{\angstrom}{\mbox{\normalfont\AA}}

\usepackage{amsmath}
\usepackage{amssymb}
\usepackage[paper=letterpaper,margin=1in]{geometry}
\usepackage{braket}
\usepackage{upgreek}

\newcommand{\beq}{\begin{eqnarray}}
\newcommand{\eeq}{\end{eqnarray}}

\usepackage{tikz}

\newtheorem{thm}{Theorem}[section]

\newtheorem{preremark}[thm]{Remark}

\begin{document}

\title{Observation of Pines' Demon in Sr$_2$RuO$_4$}
\author{Ali A. Husain}
\affiliation{Department of Physics and Materials Research Laboratory, University of Illinois, Urbana, IL, 61801, USA}
\author{Edwin W. Huang}
\affiliation{Department of Physics and Materials Research Laboratory, University of Illinois, Urbana, IL, 61801, USA}
\author{Matteo Mitrano}
\affiliation{Department of Physics, Harvard University, Cambridge, MA, 02138, USA}
\author{Melinda S. Rak}
\affiliation{Department of Physics and Materials Research Laboratory, University of Illinois, Urbana, IL, 61801, USA}
\author{Samantha I. Rubeck}
\affiliation{Department of Physics and Materials Research Laboratory, University of Illinois, Urbana, IL, 61801, USA}
\author{Xuefei Guo}
\affiliation{Department of Physics and Materials Research Laboratory, University of Illinois, Urbana, IL, 61801, USA}
\author{Hongbin Yang}
\affiliation{Department of Chemistry and Chemical Biology, Rutgers University, Piscataway, New Jersey, 08854, USA}
\author{Chanchal Sow}
\affiliation{Department of Physics, Graduate School of Science, Kyoto University, 606-8502 Kyoto, Japan}
\author{Yoshi Maeno}
\affiliation{Department of Physics, Graduate School of Science, Kyoto University, 606-8502 Kyoto, Japan}
\author{Bruno Uchoa}
\affiliation{Department of Physics and Astronomy, University of Oklahoma, Norman, OK, 73019, USA}
\author{Tai C. Chiang}
\affiliation{Department of Physics and Materials Research Laboratory, University of Illinois, Urbana, IL, 61801, USA}
\author{Philip E. Batson}
\affiliation{Department of Chemistry and Chemical Biology, Rutgers University, Piscataway, New Jersey, 08854, USA}
\author{Philip W. Phillips}
\affiliation{Department of Physics and Materials Research Laboratory, University of Illinois, Urbana, IL, 61801, USA}
\author{Peter Abbamonte}
\affiliation{Department of Physics and Materials Research Laboratory, University of Illinois, Urbana, IL, 61801, USA}

\begin{abstract}

The characteristic excitation of a metal is its plasmon, which is a quantized collective oscillation of its electron density. In 1956, David Pines predicted
that a distinct type of plasmon, dubbed a ``demon," could exist in three-dimensional metals containing more than one species of charge carrier \cite{Pines1956}. 
Consisting of out-of-phase movement of electrons in different bands, demons are acoustic, electrically neutral, and do not couple to light, so have never been detected in an equilibrium, three-dimensional metal. 
Nevertheless, demons are believed to be critical for diverse phenomena including phase transitions in mixed-valence semimetals \cite{Varma1976},  optical properties of metal nanoparticles \cite{Akashi2014}, soundarons in Weyl semimetals \cite{Afanasiev2021}, and
high temperature superconductivity in, for example, metal hydrides \cite{IhmCohen1981,Ruvalds1981,Akashi2014,Pashitskii2022}. 
Here, we present evidence for a demon in Sr$_2$RuO$_4$ from momentum-resolved electron energy-loss spectroscopy (M-EELS). Formed of electrons in the $\beta$ and $\gamma$ bands, the demon is gapless with a room temperature velocity $v=$ \SI{1.065 \pm .12}{\times 10^5 m/s} and critical momentum $q_c=0.08$ reciprocal lattice units. Its spectral weight violates low-energy sum rules, affirming its neutral character. 
Our study confirms a 66-year old prediction and suggests that demons may be a pervasive feature of multiband metals.

\end{abstract}

\maketitle
Proposed in 1952 by Pines and Bohm \cite{PinesBohm1952}, plasmons were first observed in inelastic electron scattering experiments \cite{PlatzmanWolf1973} and were one of the first confirmed examples of collective phenomena in solids. Landau referred to plasmons as ``zero sound" stressing that they are the quantum analogue of acoustic sound in a classical gas \cite{Landau1957-1,Landau1957-2} . However, unlike ordinary sound, whose frequency tends toward zero at zero momentum, $q$ (i.e., as its wavelength approaches infinity), plasmons, except in lower dimensional systems, cost a finite energy to excite, since creating a density oscillation requires overcoming the long-ranged Coulomb interaction \cite{PinesBohm1952,PinesNozieres1973}. The plasma frequency, $\omega_p$, in ordinary metals ranges from 15 eV in Al \cite{Batson1983} to 20 eV in Cu \cite{Pines1966}. 

In 1956, Pines predicted that it was possible to create a plasmon excitation with no Coulomb energy cost \cite{Pines1956}. The new collective mode, dubbed a ``demon," arises when electrons in different bands move out of phase, thereby resulting in no net transfer of charge but a modulation in the band occupancy. A demon may be thought of as a collective mode of neutral quasiparticles whose charge has been fully screened by electrons in a separate band. 
The frequency, $\omega$, of a demon mode should scale as $\omega\propto q$, vanishing as $q \rightarrow 0$ \cite{Pines1956}. Demons are believed to be crucual for many low-energy phenomena in multiband metals, including metal-insulator transitions in mixed-valence compounds \cite{Varma1976}, optical properties of metal nanocrystals \cite{Kresin1996},  soundarons in Weyl semimetals \cite{Afanasiev2021}, and high temperature superconductivity \cite{IhmCohen1981,Ruvalds1981}, for example, in metal hydrides \cite{Pashitskii2022}.


Surprisingly, while discussed widely in the theoretical literature \cite{Pines1956,IhmCohen1981,Ehrenreich1959,Varma1976,DasSarma1981,Ruvalds1981,Ku2002}, there appears to be no experimental confirmation of a demon in a three-dimensional (3D) metal, even 66 years after its prediction. 
Acoustic plasmons have been widely studied in two-dimensional (2D) metals \cite{glinka2016,Wee2009,Bhatti1996,Diaconescu2007,Park2010}, in which conventional, single-component plasmons are gapless \cite{Allen1977}. Low-energy plasmons have also been reported in layered 3D metals at $q=\pi/d$ ($d$ being the layer spacing), mostly recently by RIXS techniques \cite{Hepting2018,Nag2020}, though these excitations disperse to $\omega_p$ at $q=0$ so are not acoustic \cite{Bozovic1990}. A demon was once reported in photoexcited GaAs, though the effect is only transient \cite{Pinczuk1981}. 
A true demon, that consists of out-of-phase movement of distinct electron fluids and remains acoustic as $q \rightarrow 0$ in a 3D system, has not yet been reported. 

What makes demons difficult to detect is their inherent charge neutrality. The out-of-phase currents of the two electron fluids exactly cancel as $q \rightarrow 0$, extinguishing the long-ranged part of the Coulomb interaction. For this reason, a demon has no signature in the dielectric function of a metal, $\epsilon(q,\omega)$, in the limit of small $q$, and does not couple to light. The most promising way to detect a demon is to measure the excitations of a multiband metal at nonzero $q$, where a demon modulates the density and may be experimentally observable using EELS techniques that observed plasmons originally \cite{PlatzmanWolf1973}.

The metal we investigate is Sr$_2$RuO$_4$, which has three nested bands, $\alpha$, $\beta$, and $\gamma$, crossing the Fermi energy (Fig. 1(a)) \cite{Damascelli2001,Tamai2019}. At temperature $T \lesssim 40$ K, Sr$_2$RuO$_4$ is a good Fermi liquid showing resistivity $\rho(T) \sim T^2$,  well-defined quantum oscillations \cite{Mackenzie2003}, and the expected scattering rate in optics \cite{Stricker2014}. At higher temperature, $T \gtrsim 600$ K, Sr$_2$RuO$_4$ crosses over into a strongly interacting ``strange metal" phase in which the quasiparticles are highly damped \cite{Wang2014}, the resistivity $\rho \sim T$ and violates the Mott-Ioffe-Regel limit \cite{Tyler1998}. The strong interactions arise from Hund's coupling and are described well by dynamical mean field theory (DMFT) \cite{Tamai2019,deMedici2011}. 

As a multiband metal, Sr$_2$RuO$_4$ is a candidate for exhibiting a demon. In particular, the $\beta$ and $\gamma$ bands have quite different velocities and curvature \cite{Damascelli2001,Shen2001,Tamai2019}, reminiscent of Pines’ original conceptualization of a demon as a mode in which light electrons screen the Coulomb interaction between heavy electrons \cite{Pines1956}. Understanding whether a demon is expected in Sr$_2$RuO$_4$ requires a microscopic calculation.

We calculated the collective charge excitations of Sr$_2$RuO$_4$ by computing its dynamic charge susceptibility, $\chi(q,\omega)$, in the random phase approximation (RPA) \cite{Pines1966,PinesNozieres1973} (Supplementary Information Section II). Sr$_2$RuO$_4$ is a Fermi liquid at low energy, so we expect RPA to be a reasonable approximation for $\omega \lesssim k_B (600K) = 50$ meV. We first computed the Lindhard function using a tight binding parameterization of the energy bands \cite{Zabolotnyy2013}, and then determined the susceptibility, $\chi(q,\omega)$, using the Coulomb interaction $V(q) = e^2/\epsilon_\infty q^2$ with $\epsilon_\infty=2.3$ taken from Ref. \cite{Stricker2014}. The calculation has no adjustable parameters and no fine tuning or fitting to experimental data was done. 

Fig. 1 shows the imaginary part, $\chi''(q,\omega)$ along the (1,0,0) direction as a function of momentum, $q$, and energy, $\omega$. The most prominent feature is a sharp plasmon at $\omega_p=1.6$ eV (Fig. 2(a)), which is similar to the measured zero crossing of the real part of $\epsilon(\omega)$ in optics \cite{Stricker2014}. The plasmon exhibits a downward dispersion, which is a band structure effect similar to that observed in transition metal dichalcogenides \cite{vanWezel2011}. Note that the {\it intensity} of the plasmon (color scale) scales like $q^2$ at small momenta (Fig. 2(a)), which is consistent with the $f$-sum rule \cite{PinesNozieres1973}. This permits $\epsilon(q,0) = 1/[1+V(q) \chi(q,0) ]$ to diverge at small values of $q$, which is required in a metal in which the electric field should be completely screened over long distances.  

Looking at low energy, we see that the calculation also shows an acoustic mode (Fig. 2(b)). Its velocity,  $v=$ \SI{0.639}{eV \cdot \angstrom}, lies between the velocities of the $\beta$ and $\gamma$ bands, which is an expected property of a demon \cite{Pines1956}. Unlike the plasmon, the intensity of this excitation scales as $q^4$ (Figs. 2(b), S14), which is faster than would be expected from the $f$-sum rule. Were this the only excitation present in the material, it would imply that $\epsilon(q,0) = 1/[1+V(q) \chi(q,0) ] \rightarrow 1$ in the limit of small $q$, meaning this excitation is neutral and does not contribute to screening over large distances.

This excitation is definitively identified as a demon by examining the partial susceptibilities, $\chi_{a,b}$, which describe the linear response of the density of electrons in band $a$ due to an external potential that couples only to electrons in band $b$. As explained in Supplementary Information Section II.E,
the relative sign of $\chi''_{a,b}$ and $\chi''_{a,a}$ indicates whether electrons in the bands $a$ and $b$ oscillate in-phase. For example, if we consider the plasmon (Figure 2(c), S14), the quantities $\chi''_{\gamma,\gamma}$, $\chi''_{\beta,\beta}$, and $\chi''_{\gamma,\beta}$ are all negative, meaning the $\beta$ and $\gamma$ subbands oscillate in-phase, regardless of which is excited. The situation is different for the acoustic mode. While $\chi''_{\gamma,\gamma}$ and $\chi''_{\beta,\beta}$ are both negative (Figs. 2(d), S14), the off-diagonal term $\chi''_{\gamma,\beta}$ is positive (Fig. 2(f)), meaning that if one drives the $\gamma$ electrons, the $\beta$ electrons respond 180$^\circ$ out-of-phase. This demonstrates that the acoustic mode predicted in RPA is a true demon in that it consists of an out-of-phase oscillation between $\beta$ and $\gamma$ electrons (Fig. 1(b)). That this mode violates the $f$-sum rule indicates that a demon can never exist independent of a conventional, high energy plasmon that accomplishes the metallic screening.  

We now compare the RPA results to momentum-resolved electron energy-loss spectroscopy (M-EELS) \cite{Vig2017} measurements of the collective excitations of Sr$_2$RuO$_4$ with an energy resolution $\Delta \omega = 6$ meV and momentum resolution $\Delta q =$ \SI{0.03}{\angstrom^{-1}}. M-EELS is done in reflection mode and measures both surface and bulk excitations at nonzero momentum transfer, $q$ \cite{Vig2017}, where the signature of a demon should be clearest (Fig. 2(b)). 
Sr$_2$RuO$_4$ crystals were grown as described previously \cite{Maeno2005} and cleaved {\it in situ} in ultrahigh vacuum to reveal pristine surfaces. The surfaces were passivated by exposing to residual CO gas, which disorders the $\sqrt{2}\times \sqrt{2}$ surface reconstruction \cite{Tamai2019} and terminates surface dangling bonds \cite{Stoger2014,Tamai2019}. This treatment eliminates the surface state that complicated interpretation of early ARPES experiments \cite{Shen2001,Damascelli2001}, and results in bulk-like properties in surface measurements \cite{Tamai2019}.

M-EELS spectra at $T=300$ K at large energy transfer show a broad plasmon peak at approximately 1.2 eV (Fig. 3(b), top curve). Its width at $q = 0.12$ r.l.u. is $\sim 10^2 \times$ larger than the predicted width of the 1.6 eV plasmon in RPA. This discrepancy is  unsurprising since Sr$_2$RuO$_4$ is a non-Fermi liquid (NFL) at $\omega \gtrsim$ 50 meV \cite{Stricker2014,deMedici2011,Tamai2019,Wang2014,Tyler1998} and RPA neglects many interaction effects that could shift and damp the plasmon. Nevertheless, RPA correctly predicts its existence and approximate energy. At larger momenta, $q \geq 0.28$ r.l.u, the plasmon evolves into a featureless, energy-independent continuum similar to that observed in Bi$_2$Sr$_2$CaCu$_2$O$_{8+x}$ (Bi-2212) \cite{mitrano2018,husain2019}, though the cutoff energy in Sr$_2$RuO$_4$ is higher (1.2 eV compared to 1.0 eV in Bi-2212). This observation was confirmed by bulk, transmission EELS measurements using a Nion UltraSTEM (Supplementary Information Section I.C), establishing it as a bulk effect, and suggests that this continuum may be a generic feature of the $q \neq 0$ density response of strange metals. 

In the low-energy regime, where RPA 
predictions should be more quantitative, M-EELS reveals an acoustic mode (Fig. 4). Its energy gap at $q=0$ is less than 8 meV, an upper bound set by the tails of the elastic line (Supplementary Information Section I.H). The dispersion of the mode in the (1,0) direction is linear over most of its range, with room-temperature group velocity $v_g=$ \SI{0.701 \pm 0.082}{eV \cdot \angstrom}
( = \SI{1.065 \pm .12}{\times 10^5 m/s} ). 
At small momentum, $q<0.03$ r.l.u., the dispersion shows a quadratic ``foot," in which $E(q) \sim q^2$, which is a real effect not caused by the finite $q$ resolution of the measurement.  
The linewidth of the mode increases with increasing $q$, its FWHM rising from 7.6(3.8) meV at $q=0.03$ r.l.u. (the lowest $q$ at which it can be estimated) to 46.2(3.9) meV at $q=0.08$ r.l.u. (Fig. S8). The mode is overdamped for momenta greater than $q_c = 0.08$ r.l.u., which we identify as its critical momentum. The velocity is temperature-dependent, falling to \SI{0.485 \pm 0.081}{eV\cdot \angstrom} at $T=$ 30 K (Fig. 4(a)-(c)), and anisotropic, increasing to \SI{0.815 \pm 0.135}{eV\cdot \angstrom} in the (1,1) direction (Fig. 4(c)). 

This excitation is clearly electronic. Its velocity is $\sim 100 \times$ that of the acoustic phonons, which propagate at the sound velocity, \SI{0.008}{eV\cdot \angstrom} \cite{Braden2007}. However, it cannot be a surface plasmon, which in EELS measurements are observed at the Ritchie frequency \cite{Plummer1995}, which is 1.4 eV in Sr$_2$RuO$_4$ \cite{Stricker2014}. 
The mode velocity is, however, within 10\% of the velocity of the gapless mode predicted by RPA (Fig. 2(b)-(d), Fig. 4 (a)-(b)). We posit that this excitation is a demon, predicted by Pines 66 years ago but not seen in a 3D metal until now. 

To check this assignment, we assess whether the mode is neutral by examining the momentum dependence of its intensity. As illustrated in Fig. 2(a), the intensity of a conventional plasmon should have the same momentum dependence as the $f$-sum rule. If the excitation is neutral, its intensity should scale with a {\it higher} power, assuring $\epsilon(q,0) = 1/[1+V(q) \chi(q,0) ] \rightarrow 1$ as $q \rightarrow 0$, meaning the excitation does not contribute to screening over large distances. One complication is M-EELS measures a surface response function \cite{Vig2017}, which satisfies a different sum rule than the Lindhard susceptibility computed in RPA. It is therefore crucial that we compare to the correct sum rule for our experiment.

The $f$-sum rule for surface M-EELS is derived in Supplementary Information Section III. The result for a gapless mode is

\begin{equation}
    I_0(q) = \frac{\hbar \sigma_0 e^2 \rho_0}{m \epsilon_0 \gamma} \frac{1}{q^5},
\end{equation}

\noindent where $q$ is the momentum and $I_0(q)$ is the energy-integrated intensity of the acoustic mode (the other constants are described in the Supplementary Information). If the mode is neutral, its intensity should exhibit a power law that is {\it higher} than $q^{-5}$. The experimental intensity for the acoustic mode is shown in Fig. 4(d). The best fit gives a power law $I_0(q) \sim q^{-1.83}$. This number is larger than $-5$, indicating that the excitation is neutral. 
We conclude that this gapless excitation is Pines' demon, predicted in 1956 but not observed in a 3D material until now. 

While only now observed in Sr$_2$RuO$_4$, demons should not be rare. Any material with more than one Fermi surface could, in principle, exhibit a demon, which should be considered a new category of collective excitation in 3D solids. A demon can be thought of as a collective mode of fully screened, neutral quasiparticles or, equivalently, as a plasmon-like modulation of two different bands that, excited out-of-phase, leaves the total density uniform (Fig. 1(b)). Demons may play an important role in the low-energy physics of multiband metals more generally \cite{Varma1976,IhmCohen1981,Ruvalds1981,Akashi2014,Pashitskii2022,Kresin1996,Afanasiev2021}. 

One outstanding question is the physical origin of the $q^2$ dispersion ``foot" at $q < 0.03$ r.l.u. (Fig. 4(a)-(c)). Not observed in Fig. 2, it must arise from some effect not accounted for in RPA. Two possibilities are disorder, which is known to pin collective excitations at $q \neq 0$ \cite{Kivelson2003}, and beyond-RPA many body phenomena such as local field or excitonic effects. Further study will shed light on this matter. 

{\bf Acknowledgements.} 
We acknowledge J. Zaanen, D. van der Marel, A. Georges, M. Zingl, H. Strand, and P. Coleman for valuable discussions. 
This work was supported by the Center for Quantum Sensing and Quantum Materials, a DOE Energy Frontier Research Center, under award DE-SC0021238.
Y.M. acknowledges support from JSPS Kakenhi (Grants JP15H5852, JP15K21717 and JP17H06136) and the JSPS-EPSRC Core-to-Core Program.
B. U. acknowledges support from NSF grant DMR-2024864
Derivation of the sum rule (B.U.) was supported by NSF grant DMR-2024864.
P.A. gratefully acknowledges support from the EPiQS program of the Gordon and Betty Moore Foundation, grant GBMF9452. M.M. acknowledges support from the Alexander von Humboldt foundation.

\begin{figure}[htb]
\centering
\includegraphics[scale=1.0]{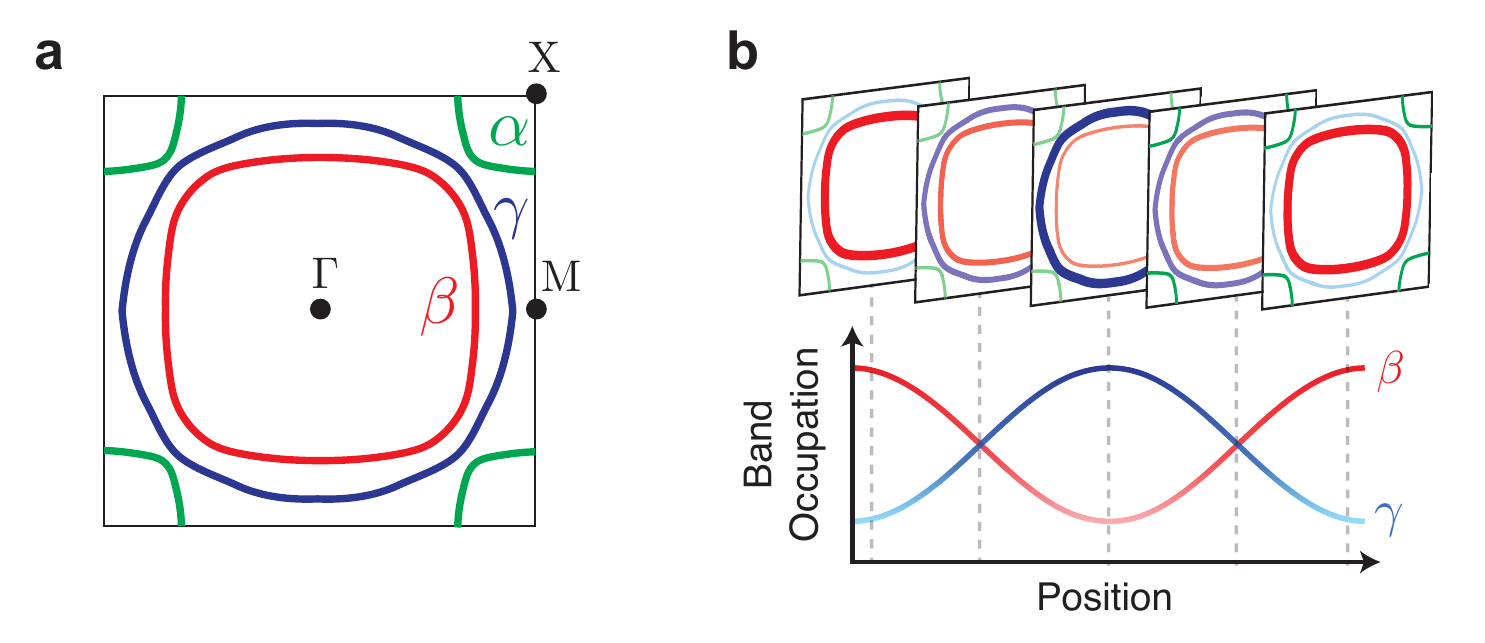}
\caption{{\bf Conceptual illustration of the demon excitation in Sr$_2$RuO$_4$.} {\bf a}, Fermi surface showing the three species of electrons, $\alpha$, $\beta$, and $\gamma$. {\bf b}, Conceptual illustration of the demon in Sr$_2$RuO$_4$, which is a modulation in the $\gamma$ and $\beta$ band fillings that keeps the overall electron density constant.}
\label{fig:1}
\end{figure}

\begin{figure}[htb]
\centering
\includegraphics{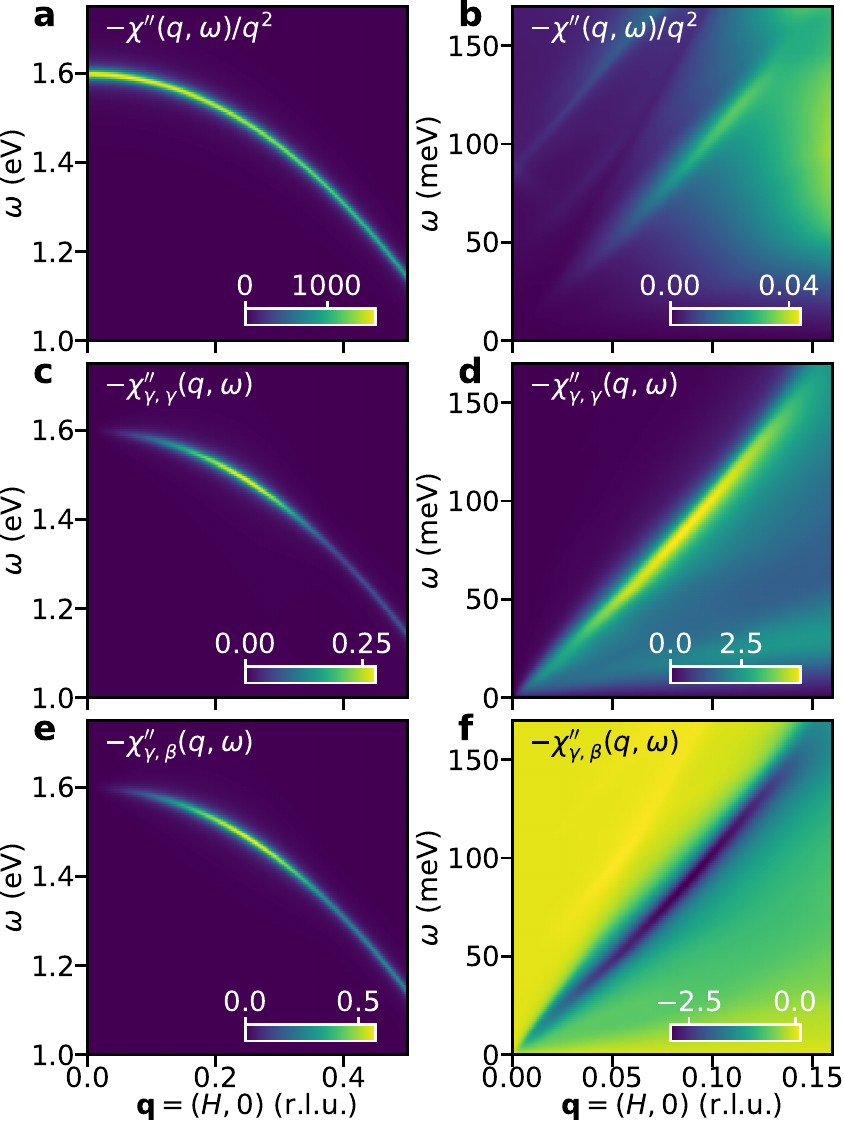}
\caption{{\bf Charge susceptibility of Sr$_2$RuO$_4$ from RPA.} {\bf a,b}, Imaginary part of the charge susceptibility $\chi(q, \omega)$, illustrating the plasmon in {\bf a} and the linearly dispersing demon in {\bf b}. {\bf c-f}, Components of the band decomposition of the susceptibility, demonstrating the in-phase response of  $\gamma$ and $\beta$ band electrons in the plasmon ({\bf c}, {\bf e}) and the out-of-phase response that gives rise to the demon ({\bf d}, {\bf f}).}
\label{fig:rpa}
\end{figure}

\begin{figure}[htb]
\centering
\includegraphics{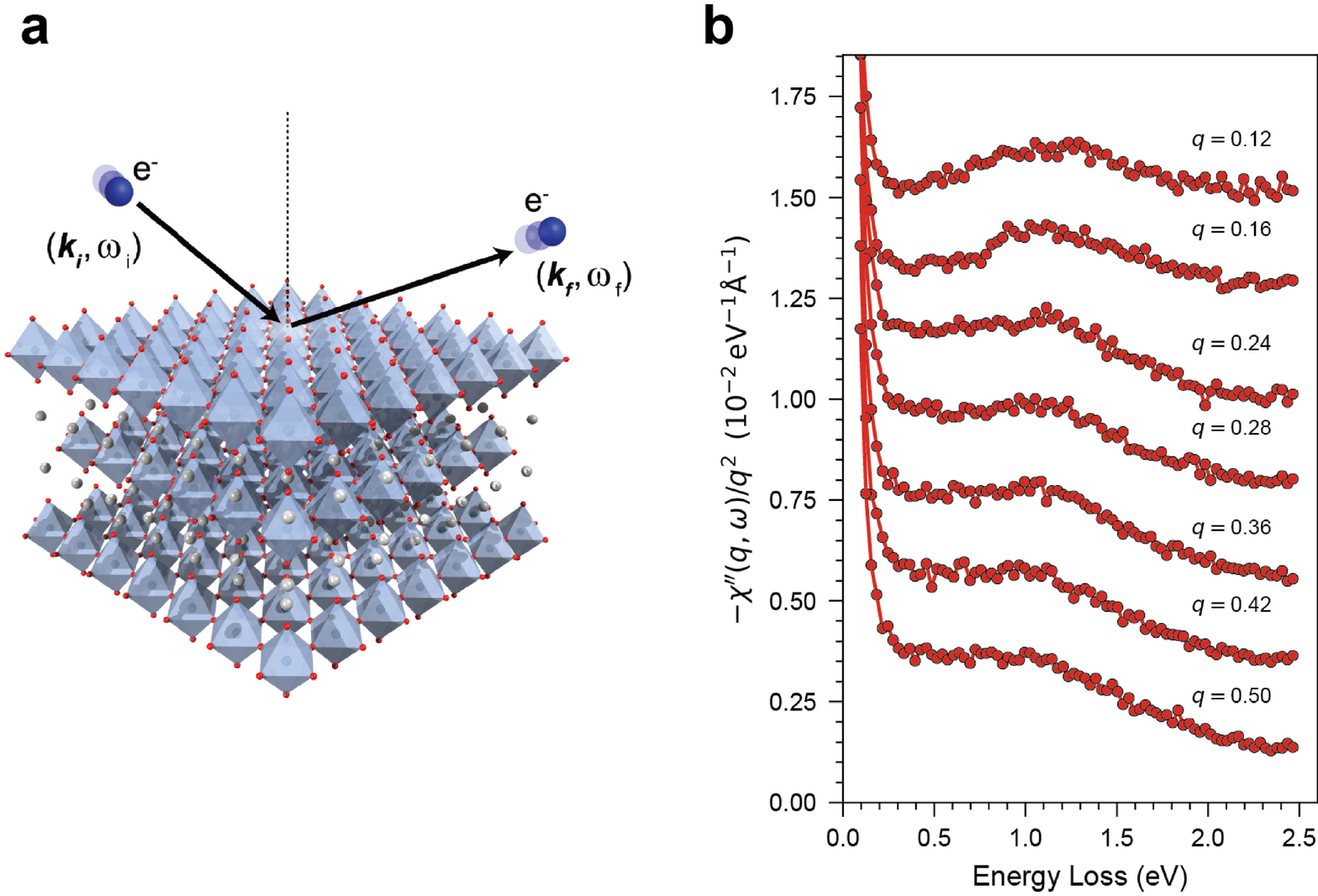}
\caption{{\bf High-energy M-EELS spectra from Sr$_2$RuO$_4$.} {\bf a} Conceptual illustration of surface M-EELS experiments from a cleaved Sr$_2$RuO$_4$ surface. {\bf b}, Fixed-$q$ (in r.l.u.) energy-loss scans for a selection of $q$ values along the (1,0) crystallographic direction, taken at $T=300$ K.
These spectra were obtained by dividing the M-EELS matrix elements and scaling the curves as described in Ref. \cite{mitrano2018}. At small momenta ($q <$ 0.16 r.l.u.), the spectra show a broad plasmon feature that peaks at 1.2 eV. At larger momenta, the data show an energy-independent continuum as was observed previously in Bi$_2$Sr$_2$CaCu$_2$O$_{8+x}$ \cite{mitrano2018}.
}
\label{fig:3}
\end{figure}

\begin{figure}[htb]
\centering
\includegraphics{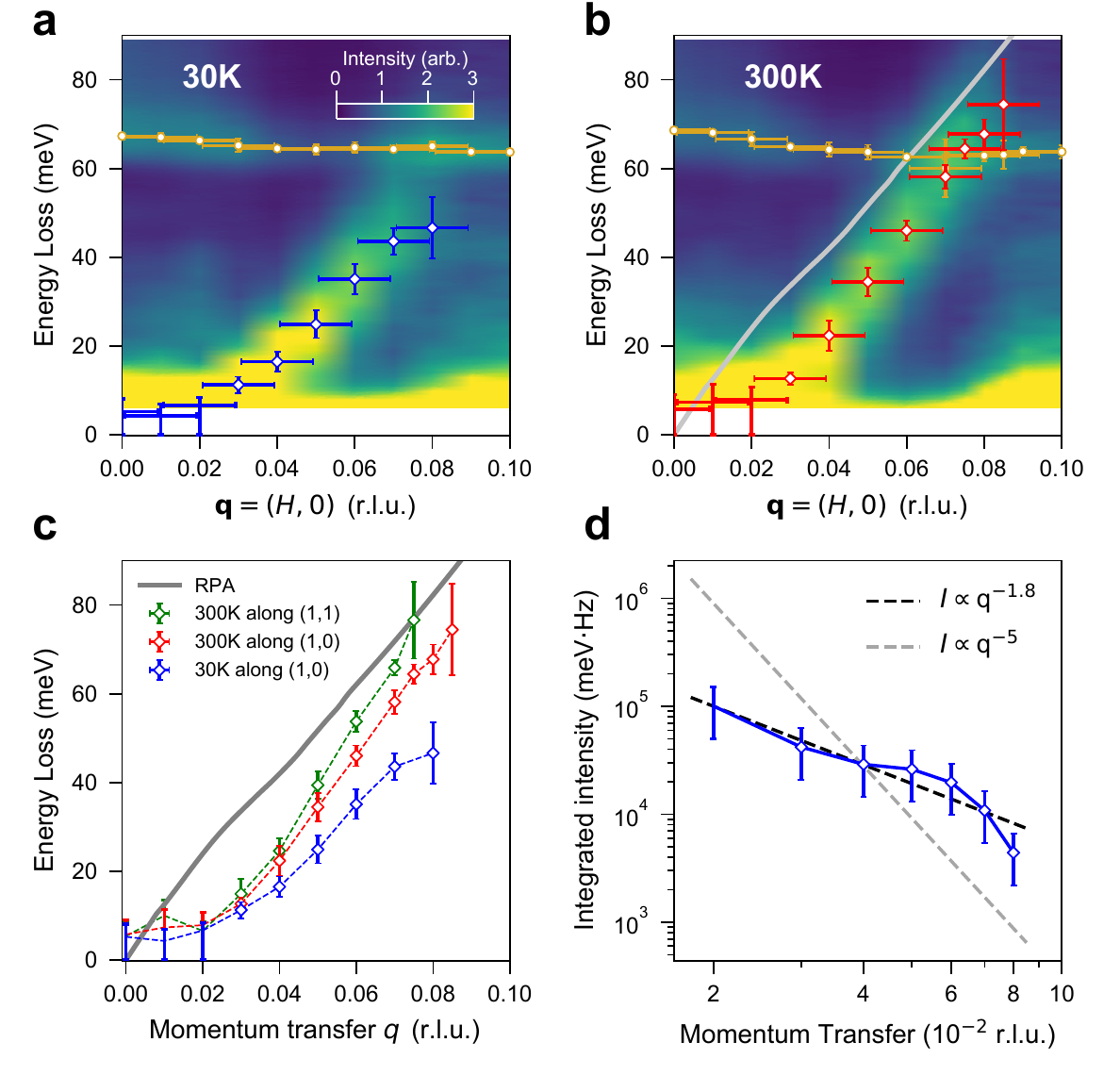}
\caption{{\bf Properties of the demon excitation in Sr$_2$RuO$_4$.} 
{\bf a,b}, Dispersion of the demon mode at 30 K (blue) and 300 K (red), compared to the predicted dispersion from RPA (grey).
The weakly dispersing excitation at 63 meV is an optical phonon. 
Vertical error bars represent the fit error, while horizontal error bars represent the momentum resolution of the instrument (see Supplementary Information).
{\bf c}, Anisotropy and temperature dependence of the demon dispersion.
Horizontal error bars are omitted from this panel for clarity.
{\bf d}, Integrated intensity of the demon excitation at $T=30$ K as a function of $q$, showing an approximate  power law
$I_0(q) \sim q^{-1.8}$ (black dashed line),
demonstrating that the excitation is neutral in the long wavelength limit.
For reference, the power law scaling expected from an ordinary excitation of
$I_0(q) \sim q^{-5}$ is shown
. 
}
\label{fig:4}
\end{figure}

\bibliography{demonbib}

\begin{thebibliography}{46}%
\makeatletter
\providecommand \@ifxundefined [1]{%
 \@ifx{#1\undefined}
}%
\providecommand \@ifnum [1]{%
 \ifnum #1\expandafter \@firstoftwo
 \else \expandafter \@secondoftwo
 \fi
}%
\providecommand \@ifx [1]{%
 \ifx #1\expandafter \@firstoftwo
 \else \expandafter \@secondoftwo
 \fi
}%
\providecommand \natexlab [1]{#1}%
\providecommand \enquote  [1]{``#1''}%
\providecommand \bibnamefont  [1]{#1}%
\providecommand \bibfnamefont [1]{#1}%
\providecommand \citenamefont [1]{#1}%
\providecommand \href@noop [0]{\@secondoftwo}%
\providecommand \href [0]{\begingroup \@sanitize@url \@href}%
\providecommand \@href[1]{\@@startlink{#1}\@@href}%
\providecommand \@@href[1]{\endgroup#1\@@endlink}%
\providecommand \@sanitize@url [0]{\catcode `\\12\catcode `\$12\catcode
  `\&12\catcode `\#12\catcode `\^12\catcode `\_12\catcode `\%12\relax}%
\providecommand \@@startlink[1]{}%
\providecommand \@@endlink[0]{}%
\providecommand \url  [0]{\begingroup\@sanitize@url \@url }%
\providecommand \@url [1]{\endgroup\@href {#1}{\urlprefix }}%
\providecommand \urlprefix  [0]{URL }%
\providecommand \Eprint [0]{\href }%
\providecommand \doibase [0]{https://doi.org/}%
\providecommand \selectlanguage [0]{\@gobble}%
\providecommand \bibinfo  [0]{\@secondoftwo}%
\providecommand \bibfield  [0]{\@secondoftwo}%
\providecommand \translation [1]{[#1]}%
\providecommand \BibitemOpen [0]{}%
\providecommand \bibitemStop [0]{}%
\providecommand \bibitemNoStop [0]{.\EOS\space}%
\providecommand \EOS [0]{\spacefactor3000\relax}%
\providecommand \BibitemShut  [1]{\csname bibitem#1\endcsname}%
\let\auto@bib@innerbib\@empty
\bibitem [{\citenamefont {Pines}(1956)}]{Pines1956}%
  \BibitemOpen
  \bibfield  {author} {\bibinfo {author} {\bibfnamefont {D.}~\bibnamefont
  {Pines}},\ }\bibfield  {title} {\bibinfo {title} {Electron interaction in
  solids},\ }\href {https://doi.org/https://doi.org/10.1139/p56-154} {\bibfield
   {journal} {\bibinfo  {journal} {Can. J. Phys.}\ }\textbf {\bibinfo {volume}
  {34}},\ \bibinfo {pages} {1379} (\bibinfo {year} {1956})}\BibitemShut
  {NoStop}%
\bibitem [{\citenamefont {Varma}(1976)}]{Varma1976}%
  \BibitemOpen
  \bibfield  {author} {\bibinfo {author} {\bibfnamefont {C.~M.}\ \bibnamefont
  {Varma}},\ }\bibfield  {title} {\bibinfo {title} {Mixed-valence compounds},\
  }\href {https://doi.org/10.1103/RevModPhys.48.219} {\bibfield  {journal}
  {\bibinfo  {journal} {Rev. Mod. Phys.}\ }\textbf {\bibinfo {volume} {48}},\
  \bibinfo {pages} {219} (\bibinfo {year} {1976})}\BibitemShut {NoStop}%
\bibitem [{\citenamefont {Akashi}\ and\ \citenamefont
  {Arita}(2014)}]{Akashi2014}%
  \BibitemOpen
  \bibfield  {author} {\bibinfo {author} {\bibfnamefont {R.}~\bibnamefont
  {Akashi}}\ and\ \bibinfo {author} {\bibfnamefont {R.}~\bibnamefont {Arita}},\
  }\bibfield  {title} {\bibinfo {title} {Density functional theory for
  plasmon-assisted superconductivity},\ }\href
  {https://doi.org/10.7566/JPSJ.83.061016} {\bibfield  {journal} {\bibinfo
  {journal} {Journal of the Physical Society of Japan}\ }\textbf {\bibinfo
  {volume} {83}},\ \bibinfo {pages} {061016} (\bibinfo {year} {2014})},\
  \Eprint {https://arxiv.org/abs/https://doi.org/10.7566/JPSJ.83.061016}
  {https://doi.org/10.7566/JPSJ.83.061016} \BibitemShut {NoStop}%
\bibitem [{\citenamefont {Afanasiev}\ \emph {et~al.}(2021)\citenamefont
  {Afanasiev}, \citenamefont {Greshnov},\ and\ \citenamefont
  {Svintsov}}]{Afanasiev2021}%
  \BibitemOpen
  \bibfield  {author} {\bibinfo {author} {\bibfnamefont {A.~N.}\ \bibnamefont
  {Afanasiev}}, \bibinfo {author} {\bibfnamefont {A.~A.}\ \bibnamefont
  {Greshnov}},\ and\ \bibinfo {author} {\bibfnamefont {D.}~\bibnamefont
  {Svintsov}},\ }\bibfield  {title} {\bibinfo {title} {Acoustic plasmons in
  type-i weyl semimetals},\ }\href
  {https://doi.org/10.1103/PhysRevB.103.205201} {\bibfield  {journal} {\bibinfo
   {journal} {Phys. Rev. B}\ }\textbf {\bibinfo {volume} {103}},\ \bibinfo
  {pages} {205201} (\bibinfo {year} {2021})}\BibitemShut {NoStop}%
\bibitem [{\citenamefont {Ihm}\ \emph {et~al.}(1981)\citenamefont {Ihm},
  \citenamefont {Cohen},\ and\ \citenamefont {Tuan}}]{IhmCohen1981}%
  \BibitemOpen
  \bibfield  {author} {\bibinfo {author} {\bibfnamefont {J.}~\bibnamefont
  {Ihm}}, \bibinfo {author} {\bibfnamefont {M.~L.}\ \bibnamefont {Cohen}},\
  and\ \bibinfo {author} {\bibfnamefont {S.~F.}\ \bibnamefont {Tuan}},\
  }\bibfield  {title} {\bibinfo {title} {Demons and superconductivity},\ }\href
  {https://doi.org/10.1103/PhysRevB.23.3258} {\bibfield  {journal} {\bibinfo
  {journal} {Phys. Rev. B}\ }\textbf {\bibinfo {volume} {23}},\ \bibinfo
  {pages} {3258} (\bibinfo {year} {1981})}\BibitemShut {NoStop}%
\bibitem [{\citenamefont {Ruvalds}(1981)}]{Ruvalds1981}%
  \BibitemOpen
  \bibfield  {author} {\bibinfo {author} {\bibfnamefont {J.}~\bibnamefont
  {Ruvalds}},\ }\bibfield  {title} {\bibinfo {title} {Are there acoustic
  plasmons?},\ }\href {https://doi.org/10.1080/00018738100101427} {\bibfield
  {journal} {\bibinfo  {journal} {Advances in Physics}\ }\textbf {\bibinfo
  {volume} {30}},\ \bibinfo {pages} {677} (\bibinfo {year} {1981})}\BibitemShut
  {NoStop}%
\bibitem [{\citenamefont {Pashitskii}\ \emph {et~al.}(2022)\citenamefont
  {Pashitskii}, \citenamefont {Pentegov},\ and\ \citenamefont
  {Semenov}}]{Pashitskii2022}%
  \BibitemOpen
  \bibfield  {author} {\bibinfo {author} {\bibfnamefont {E.~A.}\ \bibnamefont
  {Pashitskii}}, \bibinfo {author} {\bibfnamefont {V.~I.}\ \bibnamefont
  {Pentegov}},\ and\ \bibinfo {author} {\bibfnamefont {A.~V.}\ \bibnamefont
  {Semenov}},\ }\bibfield  {title} {\bibinfo {title} {Possibility for the
  anisotropic acoustic plasmons in lah10 and their role in enhancement of the
  critical temperature of superconducting transition},\ }\href
  {https://doi.org/10.1063/10.0008960} {\bibfield  {journal} {\bibinfo
  {journal} {Low Temperature Physics}\ }\textbf {\bibinfo {volume} {48}},\
  \bibinfo {pages} {26} (\bibinfo {year} {2022})},\ \Eprint
  {https://arxiv.org/abs/https://doi.org/10.1063/10.0008960}
  {https://doi.org/10.1063/10.0008960} \BibitemShut {NoStop}%
\bibitem [{\citenamefont {Pines}\ and\ \citenamefont
  {Bohm}(1952)}]{PinesBohm1952}%
  \BibitemOpen
  \bibfield  {author} {\bibinfo {author} {\bibfnamefont {D.}~\bibnamefont
  {Pines}}\ and\ \bibinfo {author} {\bibfnamefont {D.}~\bibnamefont {Bohm}},\
  }\bibfield  {title} {\bibinfo {title} {A collective description of electron
  interactions: Ii. collective $\mathrm{vs}$ individual particle aspects of the
  interactions},\ }\href {https://doi.org/10.1103/PhysRev.85.338} {\bibfield
  {journal} {\bibinfo  {journal} {Phys. Rev.}\ }\textbf {\bibinfo {volume}
  {85}},\ \bibinfo {pages} {338} (\bibinfo {year} {1952})}\BibitemShut
  {NoStop}%
\bibitem [{\citenamefont {Platzman}\ and\ \citenamefont
  {Wolff}(1973)}]{PlatzmanWolf1973}%
  \BibitemOpen
  \bibfield  {author} {\bibinfo {author} {\bibfnamefont {P.~M.}\ \bibnamefont
  {Platzman}}\ and\ \bibinfo {author} {\bibfnamefont {P.~A.}\ \bibnamefont
  {Wolff}},\ }\href@noop {} {\emph {\bibinfo {title} {Waves and Interactions in
  Solid State Plasmas}}}\ (\bibinfo  {publisher} {Academic Press, NY and
  London},\ \bibinfo {year} {1973})\BibitemShut {NoStop}%
\bibitem [{\citenamefont {Landau}(1957{\natexlab{a}})}]{Landau1957-1}%
  \BibitemOpen
  \bibfield  {author} {\bibinfo {author} {\bibfnamefont {L.~D.}\ \bibnamefont
  {Landau}},\ }\href@noop {} {\bibfield  {journal} {\bibinfo  {journal} {JETP
  (Sov. Phys.}\ }\textbf {\bibinfo {volume} {3}},\ \bibinfo {pages} {9201}
  (\bibinfo {year} {1957}{\natexlab{a}})}\BibitemShut {NoStop}%
\bibitem [{\citenamefont {Landau}(1957{\natexlab{b}})}]{Landau1957-2}%
  \BibitemOpen
  \bibfield  {author} {\bibinfo {author} {\bibfnamefont {L.~D.}\ \bibnamefont
  {Landau}},\ }\href@noop {} {\bibfield  {journal} {\bibinfo  {journal} {JETP
  (Sov. Phys.}\ }\textbf {\bibinfo {volume} {5}},\ \bibinfo {pages} {1011}
  (\bibinfo {year} {1957}{\natexlab{b}})}\BibitemShut {NoStop}%
\bibitem [{\citenamefont {Pines}\ and\ \citenamefont
  {Nozi\`eres}(1999)}]{PinesNozieres1973}%
  \BibitemOpen
  \bibfield  {author} {\bibinfo {author} {\bibfnamefont {D.}~\bibnamefont
  {Pines}}\ and\ \bibinfo {author} {\bibfnamefont {P.}~\bibnamefont
  {Nozi\`eres}},\ }\href@noop {} {\emph {\bibinfo {title} {The Theory of
  Quantum Liquids}}}\ (\bibinfo  {publisher} {Perseus Books, Cambridge, MA},\
  \bibinfo {year} {1999})\BibitemShut {NoStop}%
\bibitem [{\citenamefont {Batson}\ and\ \citenamefont
  {Silcox}(1983)}]{Batson1983}%
  \BibitemOpen
  \bibfield  {author} {\bibinfo {author} {\bibfnamefont {P.~E.}\ \bibnamefont
  {Batson}}\ and\ \bibinfo {author} {\bibfnamefont {J.}~\bibnamefont
  {Silcox}},\ }\bibfield  {title} {\bibinfo {title} {Experimental energy-loss
  function,
  $\mathrm{Im}[\ensuremath{-}\frac{1}{\ensuremath{\epsilon}}(q,\ensuremath{\omega})]$,
  for aluminum},\ }\href {https://doi.org/10.1103/PhysRevB.27.5224} {\bibfield
  {journal} {\bibinfo  {journal} {Phys. Rev. B}\ }\textbf {\bibinfo {volume}
  {27}},\ \bibinfo {pages} {5224} (\bibinfo {year} {1983})}\BibitemShut
  {NoStop}%
\bibitem [{\citenamefont {Pines}(1966)}]{Pines1966}%
  \BibitemOpen
  \bibfield  {author} {\bibinfo {author} {\bibfnamefont {D.}~\bibnamefont
  {Pines}},\ }\href@noop {} {\emph {\bibinfo {title} {Elementary Excitations in
  Solids}}}\ (\bibinfo  {publisher} {Perseus Books Publishing, Reading, MA},\
  \bibinfo {year} {1966})\BibitemShut {NoStop}%
\bibitem [{\citenamefont {Kresin}\ and\ \citenamefont
  {Kresin}(1996)}]{Kresin1996}%
  \BibitemOpen
  \bibfield  {author} {\bibinfo {author} {\bibfnamefont {V.~V.}\ \bibnamefont
  {Kresin}}\ and\ \bibinfo {author} {\bibfnamefont {V.~Z.}\ \bibnamefont
  {Kresin}},\ }\bibfield  {title} {\bibinfo {title} {Low-frequency
  “demon”-like excitations in small metal particles and their interaction
  with light},\ }\href {https://doi.org/10.1080/01418639608243526} {\bibfield
  {journal} {\bibinfo  {journal} {Philosophical Magazine B}\ }\textbf {\bibinfo
  {volume} {74}},\ \bibinfo {pages} {301} (\bibinfo {year} {1996})},\ \Eprint
  {https://arxiv.org/abs/https://doi.org/10.1080/01418639608243526}
  {https://doi.org/10.1080/01418639608243526} \BibitemShut {NoStop}%
\bibitem [{\citenamefont {Ehrenreich}\ and\ \citenamefont
  {Cohen}(1959)}]{Ehrenreich1959}%
  \BibitemOpen
  \bibfield  {author} {\bibinfo {author} {\bibfnamefont {H.}~\bibnamefont
  {Ehrenreich}}\ and\ \bibinfo {author} {\bibfnamefont {M.~H.}\ \bibnamefont
  {Cohen}},\ }\bibfield  {title} {\bibinfo {title} {Self-consistent field
  approach to the many-electron problem},\ }\href
  {https://doi.org/10.1103/PhysRev.115.786} {\bibfield  {journal} {\bibinfo
  {journal} {Phys. Rev.}\ }\textbf {\bibinfo {volume} {115}},\ \bibinfo {pages}
  {786} (\bibinfo {year} {1959})}\BibitemShut {NoStop}%
\bibitem [{\citenamefont {Das~Sarma}\ and\ \citenamefont
  {Madhukar}(1981)}]{DasSarma1981}%
  \BibitemOpen
  \bibfield  {author} {\bibinfo {author} {\bibfnamefont {S.}~\bibnamefont
  {Das~Sarma}}\ and\ \bibinfo {author} {\bibfnamefont {A.}~\bibnamefont
  {Madhukar}},\ }\bibfield  {title} {\bibinfo {title} {Collective modes of
  spatially separated, two-component, two-dimensional plasma in solids},\
  }\href {https://doi.org/10.1103/PhysRevB.23.805} {\bibfield  {journal}
  {\bibinfo  {journal} {Phys. Rev. B}\ }\textbf {\bibinfo {volume} {23}},\
  \bibinfo {pages} {805} (\bibinfo {year} {1981})}\BibitemShut {NoStop}%
\bibitem [{\citenamefont {Ku}\ \emph {et~al.}(2002)\citenamefont {Ku},
  \citenamefont {Pickett}, \citenamefont {Scalettar},\ and\ \citenamefont
  {Eguiluz}}]{Ku2002}%
  \BibitemOpen
  \bibfield  {author} {\bibinfo {author} {\bibfnamefont {W.}~\bibnamefont
  {Ku}}, \bibinfo {author} {\bibfnamefont {W.~E.}\ \bibnamefont {Pickett}},
  \bibinfo {author} {\bibfnamefont {R.~T.}\ \bibnamefont {Scalettar}},\ and\
  \bibinfo {author} {\bibfnamefont {A.~G.}\ \bibnamefont {Eguiluz}},\
  }\bibfield  {title} {\bibinfo {title} {Ab initio investigation of collective
  charge excitations in ${\mathrm{mgb}}_{2}$},\ }\href
  {https://doi.org/10.1103/PhysRevLett.88.057001} {\bibfield  {journal}
  {\bibinfo  {journal} {Phys. Rev. Lett.}\ }\textbf {\bibinfo {volume} {88}},\
  \bibinfo {pages} {057001} (\bibinfo {year} {2002})}\BibitemShut {NoStop}%
\bibitem [{\citenamefont {Glinka}\ \emph {et~al.}(2016)\citenamefont {Glinka},
  \citenamefont {Babakiray}, \citenamefont {Johnson}, \citenamefont {Holcomb},\
  and\ \citenamefont {Lederman}}]{glinka2016}%
  \BibitemOpen
  \bibfield  {author} {\bibinfo {author} {\bibfnamefont {Y.~D.}\ \bibnamefont
  {Glinka}}, \bibinfo {author} {\bibfnamefont {S.}~\bibnamefont {Babakiray}},
  \bibinfo {author} {\bibfnamefont {T.~A.}\ \bibnamefont {Johnson}}, \bibinfo
  {author} {\bibfnamefont {M.~B.}\ \bibnamefont {Holcomb}},\ and\ \bibinfo
  {author} {\bibfnamefont {D.}~\bibnamefont {Lederman}},\ }\bibfield  {title}
  {\bibinfo {title} {Nonlinear optical observation of coherent acoustic dirac
  plasmons in thin-film topological insulators},\ }\href
  {https://doi.org/https://doi.org/10.1038/ncomms13054} {\bibfield  {journal}
  {\bibinfo  {journal} {Nature Comm.}\ }\textbf {\bibinfo {volume} {7}},\
  \bibinfo {pages} {13054} (\bibinfo {year} {2016})}\BibitemShut {NoStop}%
\bibitem [{\citenamefont {Lu}\ \emph {et~al.}(2009)\citenamefont {Lu},
  \citenamefont {Loh}, \citenamefont {Huang}, \citenamefont {Chen},\ and\
  \citenamefont {Wee}}]{Wee2009}%
  \BibitemOpen
  \bibfield  {author} {\bibinfo {author} {\bibfnamefont {J.}~\bibnamefont
  {Lu}}, \bibinfo {author} {\bibfnamefont {K.~P.}\ \bibnamefont {Loh}},
  \bibinfo {author} {\bibfnamefont {H.}~\bibnamefont {Huang}}, \bibinfo
  {author} {\bibfnamefont {W.}~\bibnamefont {Chen}},\ and\ \bibinfo {author}
  {\bibfnamefont {A.~T.~S.}\ \bibnamefont {Wee}},\ }\bibfield  {title}
  {\bibinfo {title} {Plasmon dispersion on epitaxial graphene studied using
  high-resolution electron energy-loss spectroscopy},\ }\href
  {https://doi.org/10.1103/PhysRevB.80.113410} {\bibfield  {journal} {\bibinfo
  {journal} {Phys. Rev. B}\ }\textbf {\bibinfo {volume} {80}},\ \bibinfo
  {pages} {113410} (\bibinfo {year} {2009})}\BibitemShut {NoStop}%
\bibitem [{\citenamefont {Bhatti}\ \emph {et~al.}(1996)\citenamefont {Bhatti},
  \citenamefont {Richards}, \citenamefont {Hughes},\ and\ \citenamefont
  {Ritchie}}]{Bhatti1996}%
  \BibitemOpen
  \bibfield  {author} {\bibinfo {author} {\bibfnamefont {A.~S.}\ \bibnamefont
  {Bhatti}}, \bibinfo {author} {\bibfnamefont {D.}~\bibnamefont {Richards}},
  \bibinfo {author} {\bibfnamefont {H.~P.}\ \bibnamefont {Hughes}},\ and\
  \bibinfo {author} {\bibfnamefont {D.~A.}\ \bibnamefont {Ritchie}},\
  }\bibfield  {title} {\bibinfo {title} {Spatially resolved raman scattering
  from hot acoustic and optic plasmons},\ }\href
  {https://doi.org/10.1103/PhysRevB.53.11016} {\bibfield  {journal} {\bibinfo
  {journal} {Phys. Rev. B}\ }\textbf {\bibinfo {volume} {53}},\ \bibinfo
  {pages} {11016} (\bibinfo {year} {1996})}\BibitemShut {NoStop}%
\bibitem [{\citenamefont {Diaconescu}\ \emph {et~al.}(2007)\citenamefont
  {Diaconescu}, \citenamefont {Pohl}, \citenamefont {Vattuone}, \citenamefont
  {Savio}, \citenamefont {Hofmann}, \citenamefont {Silkin}, \citenamefont
  {Pitarke}, \citenamefont {Chulkov}, \citenamefont {Echenique}, \citenamefont
  {Far{\'i}as},\ and\ \citenamefont {Rocca}}]{Diaconescu2007}%
  \BibitemOpen
  \bibfield  {author} {\bibinfo {author} {\bibfnamefont {B.}~\bibnamefont
  {Diaconescu}}, \bibinfo {author} {\bibfnamefont {K.}~\bibnamefont {Pohl}},
  \bibinfo {author} {\bibfnamefont {L.}~\bibnamefont {Vattuone}}, \bibinfo
  {author} {\bibfnamefont {L.}~\bibnamefont {Savio}}, \bibinfo {author}
  {\bibfnamefont {P.}~\bibnamefont {Hofmann}}, \bibinfo {author} {\bibfnamefont
  {V.~M.}\ \bibnamefont {Silkin}}, \bibinfo {author} {\bibfnamefont {J.~M.}\
  \bibnamefont {Pitarke}}, \bibinfo {author} {\bibfnamefont {E.~V.}\
  \bibnamefont {Chulkov}}, \bibinfo {author} {\bibfnamefont {P.~M.}\
  \bibnamefont {Echenique}}, \bibinfo {author} {\bibfnamefont {D.}~\bibnamefont
  {Far{\'i}as}},\ and\ \bibinfo {author} {\bibfnamefont {M.}~\bibnamefont
  {Rocca}},\ }\bibfield  {title} {\bibinfo {title} {Low-energy acoustic
  plasmons at metal surfaces},\ }\href {https://doi.org/10.1038/nature05975}
  {\bibfield  {journal} {\bibinfo  {journal} {Nature}\ }\textbf {\bibinfo
  {volume} {448}},\ \bibinfo {pages} {57} (\bibinfo {year} {2007})}\BibitemShut
  {NoStop}%
\bibitem [{\citenamefont {Park}\ and\ \citenamefont {Palmer}(2010)}]{Park2010}%
  \BibitemOpen
  \bibfield  {author} {\bibinfo {author} {\bibfnamefont {S.~J.}\ \bibnamefont
  {Park}}\ and\ \bibinfo {author} {\bibfnamefont {R.~E.}\ \bibnamefont
  {Palmer}},\ }\bibfield  {title} {\bibinfo {title} {Acoustic plasmon on the
  au(111) surface},\ }\href {https://doi.org/10.1103/PhysRevLett.105.016801}
  {\bibfield  {journal} {\bibinfo  {journal} {Phys. Rev. Lett.}\ }\textbf
  {\bibinfo {volume} {105}},\ \bibinfo {pages} {016801} (\bibinfo {year}
  {2010})}\BibitemShut {NoStop}%
\bibitem [{\citenamefont {Allen}\ \emph {et~al.}(1977)\citenamefont {Allen},
  \citenamefont {Tsui},\ and\ \citenamefont {Logan}}]{Allen1977}%
  \BibitemOpen
  \bibfield  {author} {\bibinfo {author} {\bibfnamefont {S.~J.}\ \bibnamefont
  {Allen}}, \bibinfo {author} {\bibfnamefont {D.~C.}\ \bibnamefont {Tsui}},\
  and\ \bibinfo {author} {\bibfnamefont {R.~A.}\ \bibnamefont {Logan}},\
  }\bibfield  {title} {\bibinfo {title} {Observation of the two-dimensional
  plasmon in silicon inversion layers},\ }\href
  {https://doi.org/10.1103/PhysRevLett.38.980} {\bibfield  {journal} {\bibinfo
  {journal} {Phys. Rev. Lett.}\ }\textbf {\bibinfo {volume} {38}},\ \bibinfo
  {pages} {980} (\bibinfo {year} {1977})}\BibitemShut {NoStop}%
\bibitem [{\citenamefont {Hepting}\ \emph {et~al.}(2018)\citenamefont
  {Hepting}, \citenamefont {Chaix}, \citenamefont {Huang}, \citenamefont
  {Fumagalli}, \citenamefont {Peng}, \citenamefont {Moritz}, \citenamefont
  {Kummer}, \citenamefont {Brookes}, \citenamefont {Lee}, \citenamefont
  {Hashimoto}, \citenamefont {Sarkar}, \citenamefont {He}, \citenamefont
  {Rotundu}, \citenamefont {Lee}, \citenamefont {Greene}, \citenamefont
  {Braicovich}, \citenamefont {Ghiringhelli}, \citenamefont {Shen},
  \citenamefont {Devereaux},\ and\ \citenamefont {Lee}}]{Hepting2018}%
  \BibitemOpen
  \bibfield  {author} {\bibinfo {author} {\bibfnamefont {M.}~\bibnamefont
  {Hepting}}, \bibinfo {author} {\bibfnamefont {L.}~\bibnamefont {Chaix}},
  \bibinfo {author} {\bibfnamefont {E.~W.}\ \bibnamefont {Huang}}, \bibinfo
  {author} {\bibfnamefont {R.}~\bibnamefont {Fumagalli}}, \bibinfo {author}
  {\bibfnamefont {Y.~Y.}\ \bibnamefont {Peng}}, \bibinfo {author}
  {\bibfnamefont {B.}~\bibnamefont {Moritz}}, \bibinfo {author} {\bibfnamefont
  {K.}~\bibnamefont {Kummer}}, \bibinfo {author} {\bibfnamefont {N.~B.}\
  \bibnamefont {Brookes}}, \bibinfo {author} {\bibfnamefont {W.~C.}\
  \bibnamefont {Lee}}, \bibinfo {author} {\bibfnamefont {M.}~\bibnamefont
  {Hashimoto}}, \bibinfo {author} {\bibfnamefont {T.}~\bibnamefont {Sarkar}},
  \bibinfo {author} {\bibfnamefont {J.-F.}\ \bibnamefont {He}}, \bibinfo
  {author} {\bibfnamefont {C.~R.}\ \bibnamefont {Rotundu}}, \bibinfo {author}
  {\bibfnamefont {Y.~S.}\ \bibnamefont {Lee}}, \bibinfo {author} {\bibfnamefont
  {R.~L.}\ \bibnamefont {Greene}}, \bibinfo {author} {\bibfnamefont
  {L.}~\bibnamefont {Braicovich}}, \bibinfo {author} {\bibfnamefont
  {G.}~\bibnamefont {Ghiringhelli}}, \bibinfo {author} {\bibfnamefont {Z.~X.}\
  \bibnamefont {Shen}}, \bibinfo {author} {\bibfnamefont {T.~P.}\ \bibnamefont
  {Devereaux}},\ and\ \bibinfo {author} {\bibfnamefont {W.~S.}\ \bibnamefont
  {Lee}},\ }\bibfield  {title} {\bibinfo {title} {Three-dimensional collective
  charge excitations in electron-doped copper oxide superconductors},\ }\href
  {https://doi.org/https://doi.org/10.1038/s41586-018-0648-3} {\bibfield
  {journal} {\bibinfo  {journal} {Nature}\ }\textbf {\bibinfo {volume} {563}},\
  \bibinfo {pages} {374} (\bibinfo {year} {2018})}\BibitemShut {NoStop}%
\bibitem [{\citenamefont {Nag}\ \emph {et~al.}(2020)\citenamefont {Nag},
  \citenamefont {Zhu}, \citenamefont {Bejas}, \citenamefont {Li}, \citenamefont
  {Robarts}, \citenamefont {Yamase}, \citenamefont {Petsch}, \citenamefont
  {Song}, \citenamefont {Eisaki}, \citenamefont {Walters}, \citenamefont
  {Garc\'{\i}a-Fern\'andez}, \citenamefont {Greco}, \citenamefont {Hayden},\
  and\ \citenamefont {Zhou}}]{Nag2020}%
  \BibitemOpen
  \bibfield  {author} {\bibinfo {author} {\bibfnamefont {A.}~\bibnamefont
  {Nag}}, \bibinfo {author} {\bibfnamefont {M.}~\bibnamefont {Zhu}}, \bibinfo
  {author} {\bibfnamefont {M.}~\bibnamefont {Bejas}}, \bibinfo {author}
  {\bibfnamefont {J.}~\bibnamefont {Li}}, \bibinfo {author} {\bibfnamefont
  {H.~C.}\ \bibnamefont {Robarts}}, \bibinfo {author} {\bibfnamefont
  {H.}~\bibnamefont {Yamase}}, \bibinfo {author} {\bibfnamefont {A.~N.}\
  \bibnamefont {Petsch}}, \bibinfo {author} {\bibfnamefont {D.}~\bibnamefont
  {Song}}, \bibinfo {author} {\bibfnamefont {H.}~\bibnamefont {Eisaki}},
  \bibinfo {author} {\bibfnamefont {A.~C.}\ \bibnamefont {Walters}}, \bibinfo
  {author} {\bibfnamefont {M.}~\bibnamefont {Garc\'{\i}a-Fern\'andez}},
  \bibinfo {author} {\bibfnamefont {A.}~\bibnamefont {Greco}}, \bibinfo
  {author} {\bibfnamefont {S.~M.}\ \bibnamefont {Hayden}},\ and\ \bibinfo
  {author} {\bibfnamefont {K.-J.}\ \bibnamefont {Zhou}},\ }\bibfield  {title}
  {\bibinfo {title} {Detection of acoustic plasmons in hole-doped lanthanum and
  bismuth cuprate superconductors using resonant inelastic x-ray scattering},\
  }\href {https://doi.org/10.1103/PhysRevLett.125.257002} {\bibfield  {journal}
  {\bibinfo  {journal} {Phys. Rev. Lett.}\ }\textbf {\bibinfo {volume} {125}},\
  \bibinfo {pages} {257002} (\bibinfo {year} {2020})}\BibitemShut {NoStop}%
\bibitem [{\citenamefont {Bozovic}(1990)}]{Bozovic1990}%
  \BibitemOpen
  \bibfield  {author} {\bibinfo {author} {\bibfnamefont {I.}~\bibnamefont
  {Bozovic}},\ }\bibfield  {title} {\bibinfo {title} {Plasmons in cuprate
  superconductors},\ }\href {https://doi.org/10.1103/PhysRevB.42.1969}
  {\bibfield  {journal} {\bibinfo  {journal} {Phys. Rev. B}\ }\textbf {\bibinfo
  {volume} {42}},\ \bibinfo {pages} {1969} (\bibinfo {year}
  {1990})}\BibitemShut {NoStop}%
\bibitem [{\citenamefont {Pinczuk}\ \emph {et~al.}(1981)\citenamefont
  {Pinczuk}, \citenamefont {Shah},\ and\ \citenamefont {Wolff}}]{Pinczuk1981}%
  \BibitemOpen
  \bibfield  {author} {\bibinfo {author} {\bibfnamefont {A.}~\bibnamefont
  {Pinczuk}}, \bibinfo {author} {\bibfnamefont {J.}~\bibnamefont {Shah}},\ and\
  \bibinfo {author} {\bibfnamefont {P.~A.}\ \bibnamefont {Wolff}},\ }\bibfield
  {title} {\bibinfo {title} {Collective modes of photoexcited electron-hole
  plasmas in gaas},\ }\href {https://doi.org/10.1103/PhysRevLett.47.1487}
  {\bibfield  {journal} {\bibinfo  {journal} {Phys. Rev. Lett.}\ }\textbf
  {\bibinfo {volume} {47}},\ \bibinfo {pages} {1487} (\bibinfo {year}
  {1981})}\BibitemShut {NoStop}%
\bibitem [{\citenamefont {Damascelli}\ \emph {et~al.}(2001)\citenamefont
  {Damascelli}, \citenamefont {Shen}, \citenamefont {Lu}, \citenamefont
  {Armitage}, \citenamefont {Ronning}, \citenamefont {Feng}, \citenamefont
  {Kim}, \citenamefont {Shen}, \citenamefont {Kimura}, \citenamefont {Tokura},
  \citenamefont {Mao},\ and\ \citenamefont {Maeno}}]{Damascelli2001}%
  \BibitemOpen
  \bibfield  {author} {\bibinfo {author} {\bibfnamefont {A.}~\bibnamefont
  {Damascelli}}, \bibinfo {author} {\bibfnamefont {K.}~\bibnamefont {Shen}},
  \bibinfo {author} {\bibfnamefont {D.}~\bibnamefont {Lu}}, \bibinfo {author}
  {\bibfnamefont {N.}~\bibnamefont {Armitage}}, \bibinfo {author}
  {\bibfnamefont {F.}~\bibnamefont {Ronning}}, \bibinfo {author} {\bibfnamefont
  {D.}~\bibnamefont {Feng}}, \bibinfo {author} {\bibfnamefont {C.}~\bibnamefont
  {Kim}}, \bibinfo {author} {\bibfnamefont {Z.-X.}\ \bibnamefont {Shen}},
  \bibinfo {author} {\bibfnamefont {T.}~\bibnamefont {Kimura}}, \bibinfo
  {author} {\bibfnamefont {Y.}~\bibnamefont {Tokura}}, \bibinfo {author}
  {\bibfnamefont {Z.}~\bibnamefont {Mao}},\ and\ \bibinfo {author}
  {\bibfnamefont {Y.}~\bibnamefont {Maeno}},\ }\bibfield  {title} {\bibinfo
  {title} {Fermi surface of sr2ruo4 from angle resolved photoemission},\ }\href
  {https://doi.org/https://doi.org/10.1016/S0368-2048(00)00356-X} {\bibfield
  {journal} {\bibinfo  {journal} {Journal of Electron Spectroscopy and Related
  Phenomena}\ }\textbf {\bibinfo {volume} {114-116}},\ \bibinfo {pages} {641}
  (\bibinfo {year} {2001})},\ \bibinfo {note} {proceeding of the Eight
  International Conference on Electronic Spectroscopy and
  Structure,}\BibitemShut {NoStop}%
\bibitem [{\citenamefont {Tamai}\ \emph {et~al.}(2019)\citenamefont {Tamai},
  \citenamefont {Zingl}, \citenamefont {Rozbicki}, \citenamefont {Cappelli},
  \citenamefont {Ricc\`o}, \citenamefont {de~la Torre}, \citenamefont
  {McKeown~Walker}, \citenamefont {Bruno}, \citenamefont {King}, \citenamefont
  {Meevasana}, \citenamefont {Shi}, \citenamefont
  {Radovi\ifmmode~\acute{c}\else \'{c}\fi{}}, \citenamefont {Plumb},
  \citenamefont {Gibbs}, \citenamefont {Mackenzie}, \citenamefont {Berthod},
  \citenamefont {Strand}, \citenamefont {Kim}, \citenamefont {Georges},\ and\
  \citenamefont {Baumberger}}]{Tamai2019}%
  \BibitemOpen
  \bibfield  {author} {\bibinfo {author} {\bibfnamefont {A.}~\bibnamefont
  {Tamai}}, \bibinfo {author} {\bibfnamefont {M.}~\bibnamefont {Zingl}},
  \bibinfo {author} {\bibfnamefont {E.}~\bibnamefont {Rozbicki}}, \bibinfo
  {author} {\bibfnamefont {E.}~\bibnamefont {Cappelli}}, \bibinfo {author}
  {\bibfnamefont {S.}~\bibnamefont {Ricc\`o}}, \bibinfo {author} {\bibfnamefont
  {A.}~\bibnamefont {de~la Torre}}, \bibinfo {author} {\bibfnamefont
  {S.}~\bibnamefont {McKeown~Walker}}, \bibinfo {author} {\bibfnamefont
  {F.~Y.}\ \bibnamefont {Bruno}}, \bibinfo {author} {\bibfnamefont {P.~D.~C.}\
  \bibnamefont {King}}, \bibinfo {author} {\bibfnamefont {W.}~\bibnamefont
  {Meevasana}}, \bibinfo {author} {\bibfnamefont {M.}~\bibnamefont {Shi}},
  \bibinfo {author} {\bibfnamefont {M.}~\bibnamefont
  {Radovi\ifmmode~\acute{c}\else \'{c}\fi{}}}, \bibinfo {author} {\bibfnamefont
  {N.~C.}\ \bibnamefont {Plumb}}, \bibinfo {author} {\bibfnamefont {A.~S.}\
  \bibnamefont {Gibbs}}, \bibinfo {author} {\bibfnamefont {A.~P.}\ \bibnamefont
  {Mackenzie}}, \bibinfo {author} {\bibfnamefont {C.}~\bibnamefont {Berthod}},
  \bibinfo {author} {\bibfnamefont {H.~U.~R.}\ \bibnamefont {Strand}}, \bibinfo
  {author} {\bibfnamefont {M.}~\bibnamefont {Kim}}, \bibinfo {author}
  {\bibfnamefont {A.}~\bibnamefont {Georges}},\ and\ \bibinfo {author}
  {\bibfnamefont {F.}~\bibnamefont {Baumberger}},\ }\bibfield  {title}
  {\bibinfo {title} {High-resolution photoemission on
  ${\mathrm{sr}}_{2}{\mathrm{ruo}}_{4}$ reveals correlation-enhanced effective
  spin-orbit coupling and dominantly local self-energies},\ }\href
  {https://doi.org/10.1103/PhysRevX.9.021048} {\bibfield  {journal} {\bibinfo
  {journal} {Phys. Rev. X}\ }\textbf {\bibinfo {volume} {9}},\ \bibinfo {pages}
  {021048} (\bibinfo {year} {2019})}\BibitemShut {NoStop}%
\bibitem [{\citenamefont {Mackenzie}\ and\ \citenamefont
  {Maeno}(2003)}]{Mackenzie2003}%
  \BibitemOpen
  \bibfield  {author} {\bibinfo {author} {\bibfnamefont {A.~P.}\ \bibnamefont
  {Mackenzie}}\ and\ \bibinfo {author} {\bibfnamefont {Y.}~\bibnamefont
  {Maeno}},\ }\bibfield  {title} {\bibinfo {title} {The superconductivity of
  ${\mathrm{sr}}_{2}{\mathrm{ruo}}_{4}$ and the physics of spin-triplet
  pairing},\ }\href {https://doi.org/10.1103/RevModPhys.75.657} {\bibfield
  {journal} {\bibinfo  {journal} {Rev. Mod. Phys.}\ }\textbf {\bibinfo {volume}
  {75}},\ \bibinfo {pages} {657} (\bibinfo {year} {2003})}\BibitemShut
  {NoStop}%
\bibitem [{\citenamefont {Stricker}\ \emph {et~al.}(2014)\citenamefont
  {Stricker}, \citenamefont {Mravlje}, \citenamefont {Berthod}, \citenamefont
  {Fittipaldi}, \citenamefont {Vecchione}, \citenamefont {Georges},\ and\
  \citenamefont {van~der Marel}}]{Stricker2014}%
  \BibitemOpen
  \bibfield  {author} {\bibinfo {author} {\bibfnamefont {D.}~\bibnamefont
  {Stricker}}, \bibinfo {author} {\bibfnamefont {J.}~\bibnamefont {Mravlje}},
  \bibinfo {author} {\bibfnamefont {C.}~\bibnamefont {Berthod}}, \bibinfo
  {author} {\bibfnamefont {R.}~\bibnamefont {Fittipaldi}}, \bibinfo {author}
  {\bibfnamefont {A.}~\bibnamefont {Vecchione}}, \bibinfo {author}
  {\bibfnamefont {A.}~\bibnamefont {Georges}},\ and\ \bibinfo {author}
  {\bibfnamefont {D.}~\bibnamefont {van~der Marel}},\ }\bibfield  {title}
  {\bibinfo {title} {Optical response of ${\mathrm{sr}}_{2}{\mathrm{ruo}}_{4}$
  reveals universal fermi-liquid scaling and quasiparticles beyond landau
  theory},\ }\href {https://doi.org/10.1103/PhysRevLett.113.087404} {\bibfield
  {journal} {\bibinfo  {journal} {Phys. Rev. Lett.}\ }\textbf {\bibinfo
  {volume} {113}},\ \bibinfo {pages} {087404} (\bibinfo {year}
  {2014})}\BibitemShut {NoStop}%
\bibitem [{\citenamefont {Wang}\ \emph {et~al.}(2004)\citenamefont {Wang},
  \citenamefont {Yang}, \citenamefont {Sekharan}, \citenamefont {Ding},
  \citenamefont {Engelbrecht}, \citenamefont {Dai}, \citenamefont {Wang},
  \citenamefont {Kaminski}, \citenamefont {Valla}, \citenamefont {Kidd},
  \citenamefont {Fedorov},\ and\ \citenamefont {Johnson}}]{Wang2014}%
  \BibitemOpen
  \bibfield  {author} {\bibinfo {author} {\bibfnamefont {S.-C.}\ \bibnamefont
  {Wang}}, \bibinfo {author} {\bibfnamefont {H.-B.}\ \bibnamefont {Yang}},
  \bibinfo {author} {\bibfnamefont {A.~K.~P.}\ \bibnamefont {Sekharan}},
  \bibinfo {author} {\bibfnamefont {H.}~\bibnamefont {Ding}}, \bibinfo {author}
  {\bibfnamefont {J.~R.}\ \bibnamefont {Engelbrecht}}, \bibinfo {author}
  {\bibfnamefont {X.}~\bibnamefont {Dai}}, \bibinfo {author} {\bibfnamefont
  {Z.}~\bibnamefont {Wang}}, \bibinfo {author} {\bibfnamefont {A.}~\bibnamefont
  {Kaminski}}, \bibinfo {author} {\bibfnamefont {T.}~\bibnamefont {Valla}},
  \bibinfo {author} {\bibfnamefont {T.}~\bibnamefont {Kidd}}, \bibinfo {author}
  {\bibfnamefont {A.~V.}\ \bibnamefont {Fedorov}},\ and\ \bibinfo {author}
  {\bibfnamefont {P.~D.}\ \bibnamefont {Johnson}},\ }\bibfield  {title}
  {\bibinfo {title} {Quasiparticle line shape of sr$_2$ruo$_4$ and its relation
  to anisotropic transport},\ }\href
  {https://doi.org/10.1103/PhysRevLett.92.137002} {\bibfield  {journal}
  {\bibinfo  {journal} {Phys. Rev. Lett.}\ }\textbf {\bibinfo {volume} {92}},\
  \bibinfo {pages} {137002} (\bibinfo {year} {2004})}\BibitemShut {NoStop}%
\bibitem [{\citenamefont {Tyler}\ \emph {et~al.}(1998)\citenamefont {Tyler},
  \citenamefont {Mackenzie}, \citenamefont {NishiZaki},\ and\ \citenamefont
  {Maeno}}]{Tyler1998}%
  \BibitemOpen
  \bibfield  {author} {\bibinfo {author} {\bibfnamefont {A.~W.}\ \bibnamefont
  {Tyler}}, \bibinfo {author} {\bibfnamefont {A.~P.}\ \bibnamefont
  {Mackenzie}}, \bibinfo {author} {\bibfnamefont {S.}~\bibnamefont
  {NishiZaki}},\ and\ \bibinfo {author} {\bibfnamefont {Y.}~\bibnamefont
  {Maeno}},\ }\bibfield  {title} {\bibinfo {title} {High-temperature
  resistivity of ${\mathrm{sr}}_{2}{\mathrm{ruo}}_{4}:$ bad metallic transport
  in a good metal},\ }\href {https://doi.org/10.1103/PhysRevB.58.R10107}
  {\bibfield  {journal} {\bibinfo  {journal} {Phys. Rev. B}\ }\textbf {\bibinfo
  {volume} {58}},\ \bibinfo {pages} {R10107} (\bibinfo {year}
  {1998})}\BibitemShut {NoStop}%
\bibitem [{\citenamefont {de' Medici}\ \emph {et~al.}(2011)\citenamefont {de'
  Medici}, \citenamefont {Mravlje},\ and\ \citenamefont
  {Georges}}]{deMedici2011}%
  \BibitemOpen
  \bibfield  {author} {\bibinfo {author} {\bibfnamefont {L.}~\bibnamefont {de'
  Medici}}, \bibinfo {author} {\bibfnamefont {J.}~\bibnamefont {Mravlje}},\
  and\ \bibinfo {author} {\bibfnamefont {A.}~\bibnamefont {Georges}},\
  }\bibfield  {title} {\bibinfo {title} {Janus-faced influence of hund's rule
  coupling in strongly correlated materials},\ }\href
  {https://doi.org/10.1103/PhysRevLett.107.256401} {\bibfield  {journal}
  {\bibinfo  {journal} {Phys. Rev. Lett.}\ }\textbf {\bibinfo {volume} {107}},\
  \bibinfo {pages} {256401} (\bibinfo {year} {2011})}\BibitemShut {NoStop}%
\bibitem [{\citenamefont {Shen}\ \emph {et~al.}(2001)\citenamefont {Shen},
  \citenamefont {Damascelli}, \citenamefont {Lu}, \citenamefont {Armitage},
  \citenamefont {Ronning}, \citenamefont {Feng}, \citenamefont {Kim},
  \citenamefont {Shen}, \citenamefont {Singh}, \citenamefont {Mazin},
  \citenamefont {Nakatsuji}, \citenamefont {Mao}, \citenamefont {Maeno},
  \citenamefont {Kimura},\ and\ \citenamefont {Tokura}}]{Shen2001}%
  \BibitemOpen
  \bibfield  {author} {\bibinfo {author} {\bibfnamefont {K.~M.}\ \bibnamefont
  {Shen}}, \bibinfo {author} {\bibfnamefont {A.}~\bibnamefont {Damascelli}},
  \bibinfo {author} {\bibfnamefont {D.~H.}\ \bibnamefont {Lu}}, \bibinfo
  {author} {\bibfnamefont {N.~P.}\ \bibnamefont {Armitage}}, \bibinfo {author}
  {\bibfnamefont {F.}~\bibnamefont {Ronning}}, \bibinfo {author} {\bibfnamefont
  {D.~L.}\ \bibnamefont {Feng}}, \bibinfo {author} {\bibfnamefont
  {C.}~\bibnamefont {Kim}}, \bibinfo {author} {\bibfnamefont {Z.-X.}\
  \bibnamefont {Shen}}, \bibinfo {author} {\bibfnamefont {D.~J.}\ \bibnamefont
  {Singh}}, \bibinfo {author} {\bibfnamefont {I.~I.}\ \bibnamefont {Mazin}},
  \bibinfo {author} {\bibfnamefont {S.}~\bibnamefont {Nakatsuji}}, \bibinfo
  {author} {\bibfnamefont {Z.~Q.}\ \bibnamefont {Mao}}, \bibinfo {author}
  {\bibfnamefont {Y.}~\bibnamefont {Maeno}}, \bibinfo {author} {\bibfnamefont
  {T.}~\bibnamefont {Kimura}},\ and\ \bibinfo {author} {\bibfnamefont
  {Y.}~\bibnamefont {Tokura}},\ }\bibfield  {title} {\bibinfo {title} {Surface
  electronic structure of ${\mathrm{sr}}_{2}{\mathrm{ruo}}_{4}$},\ }\href
  {https://doi.org/10.1103/PhysRevB.64.180502} {\bibfield  {journal} {\bibinfo
  {journal} {Phys. Rev. B}\ }\textbf {\bibinfo {volume} {64}},\ \bibinfo
  {pages} {180502} (\bibinfo {year} {2001})}\BibitemShut {NoStop}%
\bibitem [{\citenamefont {Zabolotnyy}\ \emph {et~al.}(2013)\citenamefont
  {Zabolotnyy}, \citenamefont {Evtushinsky}, \citenamefont {Kordyuk},
  \citenamefont {Kim}, \citenamefont {Carleschi}, \citenamefont {Doyle},
  \citenamefont {Fittipaldi}, \citenamefont {Cuoco}, \citenamefont
  {Vecchione},\ and\ \citenamefont {Borisenko}}]{Zabolotnyy2013}%
  \BibitemOpen
  \bibfield  {author} {\bibinfo {author} {\bibfnamefont {V.}~\bibnamefont
  {Zabolotnyy}}, \bibinfo {author} {\bibfnamefont {D.}~\bibnamefont
  {Evtushinsky}}, \bibinfo {author} {\bibfnamefont {A.}~\bibnamefont
  {Kordyuk}}, \bibinfo {author} {\bibfnamefont {T.}~\bibnamefont {Kim}},
  \bibinfo {author} {\bibfnamefont {E.}~\bibnamefont {Carleschi}}, \bibinfo
  {author} {\bibfnamefont {B.}~\bibnamefont {Doyle}}, \bibinfo {author}
  {\bibfnamefont {R.}~\bibnamefont {Fittipaldi}}, \bibinfo {author}
  {\bibfnamefont {M.}~\bibnamefont {Cuoco}}, \bibinfo {author} {\bibfnamefont
  {A.}~\bibnamefont {Vecchione}},\ and\ \bibinfo {author} {\bibfnamefont
  {S.}~\bibnamefont {Borisenko}},\ }\bibfield  {title} {\bibinfo {title}
  {Renormalized band structure of sr2ruo4: A quasiparticle tight-binding
  approach},\ }\href
  {https://doi.org/https://doi.org/10.1016/j.elspec.2013.10.003} {\bibfield
  {journal} {\bibinfo  {journal} {Journal of Electron Spectroscopy and Related
  Phenomena}\ }\textbf {\bibinfo {volume} {191}},\ \bibinfo {pages} {48}
  (\bibinfo {year} {2013})}\BibitemShut {NoStop}%
\bibitem [{\citenamefont {van Wezel}\ \emph {et~al.}(2011)\citenamefont {van
  Wezel}, \citenamefont {Schuster}, \citenamefont {K\"onig}, \citenamefont
  {Knupfer}, \citenamefont {van~den Brink}, \citenamefont {Berger},\ and\
  \citenamefont {B\"uchner}}]{vanWezel2011}%
  \BibitemOpen
  \bibfield  {author} {\bibinfo {author} {\bibfnamefont {J.}~\bibnamefont {van
  Wezel}}, \bibinfo {author} {\bibfnamefont {R.}~\bibnamefont {Schuster}},
  \bibinfo {author} {\bibfnamefont {A.}~\bibnamefont {K\"onig}}, \bibinfo
  {author} {\bibfnamefont {M.}~\bibnamefont {Knupfer}}, \bibinfo {author}
  {\bibfnamefont {J.}~\bibnamefont {van~den Brink}}, \bibinfo {author}
  {\bibfnamefont {H.}~\bibnamefont {Berger}},\ and\ \bibinfo {author}
  {\bibfnamefont {B.}~\bibnamefont {B\"uchner}},\ }\bibfield  {title} {\bibinfo
  {title} {Effect of charge order on the plasmon dispersion in transition-metal
  dichalcogenides},\ }\href {https://doi.org/10.1103/PhysRevLett.107.176404}
  {\bibfield  {journal} {\bibinfo  {journal} {Phys. Rev. Lett.}\ }\textbf
  {\bibinfo {volume} {107}},\ \bibinfo {pages} {176404} (\bibinfo {year}
  {2011})}\BibitemShut {NoStop}%
\bibitem [{\citenamefont {Vig}\ \emph {et~al.}(2017)\citenamefont {Vig},
  \citenamefont {Kogar}, \citenamefont {Mitrano}, \citenamefont {Husain},
  \citenamefont {Mishra}, \citenamefont {Rak}, \citenamefont {Venema},
  \citenamefont {Johnson}, \citenamefont {Gu}, \citenamefont {Fradkin},
  \citenamefont {Norman},\ and\ \citenamefont {Abbamonte}}]{Vig2017}%
  \BibitemOpen
  \bibfield  {author} {\bibinfo {author} {\bibfnamefont {S.}~\bibnamefont
  {Vig}}, \bibinfo {author} {\bibfnamefont {A.}~\bibnamefont {Kogar}}, \bibinfo
  {author} {\bibfnamefont {M.}~\bibnamefont {Mitrano}}, \bibinfo {author}
  {\bibfnamefont {A.~A.}\ \bibnamefont {Husain}}, \bibinfo {author}
  {\bibfnamefont {V.}~\bibnamefont {Mishra}}, \bibinfo {author} {\bibfnamefont
  {M.~S.}\ \bibnamefont {Rak}}, \bibinfo {author} {\bibfnamefont
  {L.}~\bibnamefont {Venema}}, \bibinfo {author} {\bibfnamefont {P.~D.}\
  \bibnamefont {Johnson}}, \bibinfo {author} {\bibfnamefont {G.~D.}\
  \bibnamefont {Gu}}, \bibinfo {author} {\bibfnamefont {E.}~\bibnamefont
  {Fradkin}}, \bibinfo {author} {\bibfnamefont {M.~R.}\ \bibnamefont
  {Norman}},\ and\ \bibinfo {author} {\bibfnamefont {P.}~\bibnamefont
  {Abbamonte}},\ }\bibfield  {title} {\bibinfo {title} {{Measurement of the
  dynamic charge response of materials using low-energy, momentum-resolved
  electron energy-loss spectroscopy (M-EELS)}},\ }\href
  {https://doi.org/10.21468/SciPostPhys.3.4.026} {\bibfield  {journal}
  {\bibinfo  {journal} {SciPost Phys.}\ }\textbf {\bibinfo {volume} {3}},\
  \bibinfo {pages} {026} (\bibinfo {year} {2017})}\BibitemShut {NoStop}%
\bibitem [{\citenamefont {Fittipaldi}\ \emph {et~al.}(2005)\citenamefont
  {Fittipaldi}, \citenamefont {Vecchione}, \citenamefont {Fusanobori},
  \citenamefont {Takizawa}, \citenamefont {Yaguchi}, \citenamefont {Hooper},
  \citenamefont {Perry},\ and\ \citenamefont {Maeno}}]{Maeno2005}%
  \BibitemOpen
  \bibfield  {author} {\bibinfo {author} {\bibfnamefont {R.}~\bibnamefont
  {Fittipaldi}}, \bibinfo {author} {\bibfnamefont {A.}~\bibnamefont
  {Vecchione}}, \bibinfo {author} {\bibfnamefont {S.}~\bibnamefont
  {Fusanobori}}, \bibinfo {author} {\bibfnamefont {K.}~\bibnamefont
  {Takizawa}}, \bibinfo {author} {\bibfnamefont {H.}~\bibnamefont {Yaguchi}},
  \bibinfo {author} {\bibfnamefont {J.}~\bibnamefont {Hooper}}, \bibinfo
  {author} {\bibfnamefont {R.}~\bibnamefont {Perry}},\ and\ \bibinfo {author}
  {\bibfnamefont {Y.}~\bibnamefont {Maeno}},\ }\bibfield  {title} {\bibinfo
  {title} {Crystal growth of the new sr2ruo4–sr3ru2o7 eutectic system by a
  floating-zone method},\ }\href
  {https://doi.org/https://doi.org/10.1016/j.jcrysgro.2005.04.104} {\bibfield
  {journal} {\bibinfo  {journal} {Journal of Crystal Growth}\ }\textbf
  {\bibinfo {volume} {282}},\ \bibinfo {pages} {152} (\bibinfo {year}
  {2005})}\BibitemShut {NoStop}%
\bibitem [{\citenamefont {St\"oger}\ \emph {et~al.}(2014)\citenamefont
  {St\"oger}, \citenamefont {Hieckel}, \citenamefont {Mittendorfer},
  \citenamefont {Wang}, \citenamefont {Fobes}, \citenamefont {Peng},
  \citenamefont {Mao}, \citenamefont {Schmid}, \citenamefont {Redinger},\ and\
  \citenamefont {Diebold}}]{Stoger2014}%
  \BibitemOpen
  \bibfield  {author} {\bibinfo {author} {\bibfnamefont {B.}~\bibnamefont
  {St\"oger}}, \bibinfo {author} {\bibfnamefont {M.}~\bibnamefont {Hieckel}},
  \bibinfo {author} {\bibfnamefont {F.}~\bibnamefont {Mittendorfer}}, \bibinfo
  {author} {\bibfnamefont {Z.}~\bibnamefont {Wang}}, \bibinfo {author}
  {\bibfnamefont {D.}~\bibnamefont {Fobes}}, \bibinfo {author} {\bibfnamefont
  {J.}~\bibnamefont {Peng}}, \bibinfo {author} {\bibfnamefont {Z.}~\bibnamefont
  {Mao}}, \bibinfo {author} {\bibfnamefont {M.}~\bibnamefont {Schmid}},
  \bibinfo {author} {\bibfnamefont {J.}~\bibnamefont {Redinger}},\ and\
  \bibinfo {author} {\bibfnamefont {U.}~\bibnamefont {Diebold}},\ }\bibfield
  {title} {\bibinfo {title} {High chemical activity of a perovskite surface:
  Reaction of co with ${\mathrm{sr}}_{3}{\mathrm{ru}}_{2}{\mathrm{o}}_{7}$},\
  }\href {https://doi.org/10.1103/PhysRevLett.113.116101} {\bibfield  {journal}
  {\bibinfo  {journal} {Phys. Rev. Lett.}\ }\textbf {\bibinfo {volume} {113}},\
  \bibinfo {pages} {116101} (\bibinfo {year} {2014})}\BibitemShut {NoStop}%
\bibitem [{\citenamefont {Mitrano}\ \emph {et~al.}(2018)\citenamefont
  {Mitrano}, \citenamefont {Husain}, \citenamefont {Vig}, \citenamefont
  {Kogar}, \citenamefont {Rak}, \citenamefont {Rubeck}, \citenamefont
  {Schmalian}, \citenamefont {Uchoa}, \citenamefont {Schneeloch}, \citenamefont
  {Zhong}, \citenamefont {Gu},\ and\ \citenamefont {Abbamonte}}]{mitrano2018}%
  \BibitemOpen
  \bibfield  {author} {\bibinfo {author} {\bibfnamefont {M.}~\bibnamefont
  {Mitrano}}, \bibinfo {author} {\bibfnamefont {A.~A.}\ \bibnamefont {Husain}},
  \bibinfo {author} {\bibfnamefont {S.}~\bibnamefont {Vig}}, \bibinfo {author}
  {\bibfnamefont {A.}~\bibnamefont {Kogar}}, \bibinfo {author} {\bibfnamefont
  {M.~S.}\ \bibnamefont {Rak}}, \bibinfo {author} {\bibfnamefont {S.~I.}\
  \bibnamefont {Rubeck}}, \bibinfo {author} {\bibfnamefont {J.}~\bibnamefont
  {Schmalian}}, \bibinfo {author} {\bibfnamefont {B.}~\bibnamefont {Uchoa}},
  \bibinfo {author} {\bibfnamefont {J.}~\bibnamefont {Schneeloch}}, \bibinfo
  {author} {\bibfnamefont {R.}~\bibnamefont {Zhong}}, \bibinfo {author}
  {\bibfnamefont {G.~D.}\ \bibnamefont {Gu}},\ and\ \bibinfo {author}
  {\bibfnamefont {P.}~\bibnamefont {Abbamonte}},\ }\bibfield  {title} {\bibinfo
  {title} {Anomalous density fluctuations in a strange metal},\ }\href
  {https://doi.org/10.1073/pnas.1721495115} {\bibfield  {journal} {\bibinfo
  {journal} {Proceedings of the National Academy of Sciences}\ }\textbf
  {\bibinfo {volume} {115}},\ \bibinfo {pages} {5392} (\bibinfo {year}
  {2018})},\ \Eprint
  {https://arxiv.org/abs/https://www.pnas.org/doi/pdf/10.1073/pnas.1721495115}
  {https://www.pnas.org/doi/pdf/10.1073/pnas.1721495115} \BibitemShut {NoStop}%
\bibitem [{\citenamefont {Husain}\ \emph {et~al.}(2019)\citenamefont {Husain},
  \citenamefont {Mitrano}, \citenamefont {Rak}, \citenamefont {Rubeck},
  \citenamefont {Uchoa}, \citenamefont {March}, \citenamefont {Dwyer},
  \citenamefont {Schneeloch}, \citenamefont {Zhong}, \citenamefont {Gu},\ and\
  \citenamefont {Abbamonte}}]{husain2019}%
  \BibitemOpen
  \bibfield  {author} {\bibinfo {author} {\bibfnamefont {A.~A.}\ \bibnamefont
  {Husain}}, \bibinfo {author} {\bibfnamefont {M.}~\bibnamefont {Mitrano}},
  \bibinfo {author} {\bibfnamefont {M.~S.}\ \bibnamefont {Rak}}, \bibinfo
  {author} {\bibfnamefont {S.}~\bibnamefont {Rubeck}}, \bibinfo {author}
  {\bibfnamefont {B.}~\bibnamefont {Uchoa}}, \bibinfo {author} {\bibfnamefont
  {K.}~\bibnamefont {March}}, \bibinfo {author} {\bibfnamefont
  {C.}~\bibnamefont {Dwyer}}, \bibinfo {author} {\bibfnamefont
  {J.}~\bibnamefont {Schneeloch}}, \bibinfo {author} {\bibfnamefont
  {R.}~\bibnamefont {Zhong}}, \bibinfo {author} {\bibfnamefont {G.~D.}\
  \bibnamefont {Gu}},\ and\ \bibinfo {author} {\bibfnamefont {P.}~\bibnamefont
  {Abbamonte}},\ }\bibfield  {title} {\bibinfo {title} {Crossover of charge
  fluctuations across the strange metal phase diagram},\ }\href
  {https://doi.org/10.1103/PhysRevX.9.041062} {\bibfield  {journal} {\bibinfo
  {journal} {Phys. Rev. X}\ }\textbf {\bibinfo {volume} {9}},\ \bibinfo {pages}
  {041062} (\bibinfo {year} {2019})}\BibitemShut {NoStop}%
\bibitem [{\citenamefont {Braden}\ \emph {et~al.}(2007)\citenamefont {Braden},
  \citenamefont {Reichardt}, \citenamefont {Sidis}, \citenamefont {Mao},\ and\
  \citenamefont {Maeno}}]{Braden2007}%
  \BibitemOpen
  \bibfield  {author} {\bibinfo {author} {\bibfnamefont {M.}~\bibnamefont
  {Braden}}, \bibinfo {author} {\bibfnamefont {W.}~\bibnamefont {Reichardt}},
  \bibinfo {author} {\bibfnamefont {Y.}~\bibnamefont {Sidis}}, \bibinfo
  {author} {\bibfnamefont {Z.}~\bibnamefont {Mao}},\ and\ \bibinfo {author}
  {\bibfnamefont {Y.}~\bibnamefont {Maeno}},\ }\bibfield  {title} {\bibinfo
  {title} {Lattice dynamics and electron-phonon coupling in
  ${\mathrm{sr}}_{2}\mathrm{Ru}{\mathrm{o}}_{4}$: Inelastic neutron scattering
  and shell-model calculations},\ }\href
  {https://doi.org/10.1103/PhysRevB.76.014505} {\bibfield  {journal} {\bibinfo
  {journal} {Phys. Rev. B}\ }\textbf {\bibinfo {volume} {76}},\ \bibinfo
  {pages} {014505} (\bibinfo {year} {2007})}\BibitemShut {NoStop}%
\bibitem [{\citenamefont {Plummer}\ \emph {et~al.}(1995)\citenamefont
  {Plummer}, \citenamefont {Tsuei},\ and\ \citenamefont {Kim}}]{Plummer1995}%
  \BibitemOpen
  \bibfield  {author} {\bibinfo {author} {\bibfnamefont {W.}~\bibnamefont
  {Plummer}}, \bibinfo {author} {\bibfnamefont {K.-D.}\ \bibnamefont {Tsuei}},\
  and\ \bibinfo {author} {\bibfnamefont {B.-O.}\ \bibnamefont {Kim}},\
  }\bibfield  {title} {\bibinfo {title} {The impact of the concept of a surface
  plasmon},\ }\href
  {https://doi.org/https://doi.org/10.1016/0168-583X(95)00311-8} {\bibfield
  {journal} {\bibinfo  {journal} {Nuclear Instruments and Methods in Physics
  Research Section B: Beam Interactions with Materials and Atoms}\ }\textbf
  {\bibinfo {volume} {96}},\ \bibinfo {pages} {448} (\bibinfo {year}
  {1995})}\BibitemShut {NoStop}%
\bibitem [{\citenamefont {Kivelson}\ \emph {et~al.}(2003)\citenamefont
  {Kivelson}, \citenamefont {Bindloss}, \citenamefont {Fradkin}, \citenamefont
  {Oganesyan}, \citenamefont {Tranquada}, \citenamefont {Kapitulnik},\ and\
  \citenamefont {Howald}}]{Kivelson2003}%
  \BibitemOpen
  \bibfield  {author} {\bibinfo {author} {\bibfnamefont {S.~A.}\ \bibnamefont
  {Kivelson}}, \bibinfo {author} {\bibfnamefont {I.~P.}\ \bibnamefont
  {Bindloss}}, \bibinfo {author} {\bibfnamefont {E.}~\bibnamefont {Fradkin}},
  \bibinfo {author} {\bibfnamefont {V.}~\bibnamefont {Oganesyan}}, \bibinfo
  {author} {\bibfnamefont {J.~M.}\ \bibnamefont {Tranquada}}, \bibinfo {author}
  {\bibfnamefont {A.}~\bibnamefont {Kapitulnik}},\ and\ \bibinfo {author}
  {\bibfnamefont {C.}~\bibnamefont {Howald}},\ }\bibfield  {title} {\bibinfo
  {title} {How to detect fluctuating stripes in the high-temperature
  superconductors},\ }\href {https://doi.org/10.1103/RevModPhys.75.1201}
  {\bibfield  {journal} {\bibinfo  {journal} {Rev. Mod. Phys.}\ }\textbf
  {\bibinfo {volume} {75}},\ \bibinfo {pages} {1201} (\bibinfo {year}
  {2003})}\BibitemShut {NoStop}%
\end{thebibliography}%

\end{document}